\documentclass[fleqn,usenatbib,dvipsnames]{mnras}

\usepackage[T1]{fontenc}

\DeclareRobustCommand{\VAN}[3]{#2}
\let\VANthebibliography\thebibliography
\def\thebibliography{\DeclareRobustCommand{\VAN}[3]{##3}\VANthebibliography}

\usepackage{graphicx}	
\usepackage{amsmath}	
\usepackage{siunitx}    
\usepackage{rotating}
\usepackage{adjustbox}
\usepackage{multirow}
\usepackage{mathtools}
\usepackage{nccmath}
\usepackage{newtxtext,newtxmath}
\usepackage{float}
\usepackage{enumitem}
\usepackage{xcolor}

\title[Spectroscopic and Photometric evolution of SN\,2020acat]{SN\,2020acat: A `purr'-fect example of a fast rising Type IIb Supernova}

\author[Medler et al.]{K.~Medler$^{1}$ \thanks{E-mail: K.Medler@2019.ljmu.ac.uk},
P.~A.~Mazzali$^{1,2}$,
J.~Teffs $^{1}$,
C.~Ashall $^{3}$,
J.P.~Anderson $^{4}$,
I.~Arcavi $^{5,6}$,
\newauthor
S.~Benetti $^{7}$,
K.~A.~Bostroem $^{8}$
J.~Burke $^{9,10}$,
Y.-Z.~Cai $^{11}$,
P.~Charalampopoulos $^{12}$,
\newauthor
N.~Elias-Rosa $^{7,13}$,
M.~Ergon $^{14,15}$,
L.~Galbany $^{13,16}$,
M.~Gromadzki $^{17}$,
D.~Hiramatsu $^{18,19}$,
\newauthor
D.~A.~Howell $^{9,10}$,
C.~Inserra $^{20}$,
P.~Lundqvist $^{14,15}$,
C.~McCully $^{9,10}$,
T.~M\"uller-Bravo $^{13}$,
\newauthor
M.~Newsome $^{9,10}$,
M.~Nicholl $^{20}$,
E.~Padilla Gonzalez $^{9,10}$,
E.~Paraskeva $^{22,23}$,
\newauthor
A.~Pastorello $^{7}$,
C.~Pellegrino $^{9,10}$,
P.~J.~Pessi $^{24}$,
A.~Requitti $^{7,25,26}$,
T.~M.~Reynolds $^{27}$,
\newauthor
R.~Roy $^{28}$,
G.~Terreran $^{9,10}$,
L.~Tomasella $^{7}$,
D.~R.~Young $^{29}$ \\
\\
Affiliations are given after the references.
}
\date{Accepted XXX. Received YYY; in original form ZZZ}
\pubyear{2022}

\begin{document}
\label{firstpage}
\pagerange{\pageref{firstpage}--\pageref{lastpage}}
\maketitle

\begin{abstract}
The Ultra-Violet (UV) and Near Infrared (NIR) photometric and optical spectroscopic observations of SN\,2020acat covering $\sim \! \! 250$ days after explosion are presented here. Using the fast rising photometric observations, spanning from the UV to NIR wavelengths, a pseudo-bolometric light curve was constructed and compared to several other well-observed Type IIb supernovae (SNe\,IIb). SN\,2020acat displayed a very short rise time reaching a peak luminosity of $\mathrm{Log_{10}}(L) = 42.49 \pm 0.15 \, \mathrm{erg \, s^{-1}}$ in only $\sim \! \! 14.6 \pm 0.3$ days. From modelling of the pseudo-bolometric light curve, we estimated a total mass of \Nifs\ synthesised by SN\,2020acat of $0.13 \pm 0.02$ \msun, with an ejecta mass of $2.3 \pm 0.3$ \msun\ and a kinetic energy of $1.2 \pm 0.2 \times 10^{51}$ \erg. The optical spectra of SN\,2020acat display hydrogen signatures well into the transitional period ($\gtrsim 100$ days), between the photospheric and the nebular phases. The spectra also display a strong feature around $4900 \, \si{\angstrom}$ that cannot be solely accounted for by the presence of the \FeII\ $5018$ line. We suggest that the \FeII\ feature was augmented by \HeI\ $5016$ and possibly by the presence of \NII\ $5005$. From both photometric and spectroscopic analysis, we inferred that the progenitor of SN\,2020acat was an intermediate mass compact star with a \mzams\ of $18 - 22$ \msun.
\end{abstract}




\section{Introduction}

Type IIb supernovae (SNe\,IIb) belong to a subcategory of stripped envelope (SE) core-collapse supernovae (CC-SNe) that result from the explosion of stars with an initial zero age main sequence mass of \mzams $> 8$ \msun\, \citep{2009ARA&A..47...63S}. 
Prior to core-collapse, the progenitors of SE-SNe undergo the stripping of their outer envelope leaving a thin hydrogen layer, an open helium layer or a bare CO core at the time of explosion. The mechanism that removes mass from the outer envelope is thought to be either mass transfer via interaction with a companion star within a binary system, likely during the common envelope phase \citep[e.g.][]{1992ApJ...391..246P, 10.1093/mnras/stz3224}, or through the ejection of the outer layer by strong stellar winds from high metallicity stars during the Wolf-Rayet phase \citep[e.g.][]{10.1093/mnras/stv2283}. This presents multiple possible formation channels for SE-SNe, resulting from either a binary or a single star system. Yet, in recent years it seems that the binary system has become the favoured formation method for SE-SNe, as the stellar winds predicted for the single star channel would not be able to account for the current observed rate of SE-SNe \citep{10.1111/j.1365-2966.2011.17229.x}. 

The various classifications of SE-SNe arise from progenitor stars that have undergone drastically different degrees of stripping prior to collapse, and display a large variety of spectral signatures throughout their evolution. If all or the majority of hydrogen is removed from the outer envelope, a H-poor Type Ib/c SN (SN\,Ib/c) will occur. As a consequence, SNe\,Ib/c lack any prominent hydrogen features within their spectral evolution, and are dominated by helium and heavier elements. However, if there remains a significant mass of hydrogen, between $0.001 \text{ and } 0.5$ \msun\ \citep{Yoon_2017, Sravan_2019}, a SN\,IIb will occur. Due to the presence of a thin hydrogen shell prior to CC, SNe\,IIb display strong hydrogen features in their spectra during the photospheric phase, with the \Ha\ and \Hb\ lines being the most prominent features during this phase, along with several strong helium lines. The hydrogen features in SNe\,IIb fade over time until the spectra become SN\,Ib-like \citep{Filippenko:2000yf}. During the post maximum evolution, helium features dominate the spectra until the event transitions into the nebular phase, when oxygen lines become the strongest features within the spectrum, along with some iron-group elements. 
The first photometrically and spectroscopiccally well observed SNe that displayed signatures of thin hydrogen shells prior to collapse were SN\,1987K \citep{1989ApJ...343..323F} and SN\,1993J \citep{1994ApJ...429..300W}, with the latter becoming the template for the SNe\,IIb classification. 

SNe\,IIb have a similar bolometric light curve shape to SNe\,Ib, although they seem to have a lower average peak luminosity, which is powered by the decay of radioactive \Nifs\ \citep{2019MNRAS.485.1559P}, suggesting that SNe\,IIb possess a lower average mass of \Nifs\ that was synthesised during the explosion. Several SNe\,IIb and SNe\,Ib also display an initial very bright peak several days before the main \Nifs\, powered peak. The shock cooling tail of the initial bright peak was seen in SN\,1993J \citep{1993ApJ...417L..71W}, SN\,2008D \citep{Malesani_2009}, SN\,2011fu \citep{10.1093/mnras/stv1972}, SN\,2013df \citep{Morales_Garoffolo_2014} and SN\,2016gkg \citep{2018Natur.554..497B}. This initial luminous peak is thought to result from the shock breakout on the stellar surface. As such, the shock breakout phase of the bolometric light curve contains vital information on the compactness of the progenitor surface prior to core-collapse. Both the presence and lack of any observed shock-cooling phase, especially if the SNe is caught very early, can provide details on the type of progenitor. 

In this paper, we present the ultra violet (UV), optical and near infrared (NIR) observations of the Type IIb SN\,2020acat. Photometric coverage includes UV to NIR observations, while the spectra cover the optical observations. In Section \ref{Host}, we discuss the distance and reddening associated with the host galaxy of SN\,2020acat, along with the explosion date. In Section \ref{aquis}, we present the acquisition of the UV to NIR photometric data and the spectral evolution of SN\,2020acat. In Section \ref{phot}, we discuss the fast rising pseudo-bolometric light curve and the derived physical parameters of SN\,2020acat, along with comparisons with other SNe\,IIb. Then, in Section \ref{spec}, we examine the spectroscopic evolution of SN\,2020acat within the photospheric, transitional and early nebular phases, including the line velocity of the \Ha, \Hb, \HeI\ $5867$ and \FeII\ $5018$. We also compare the optical spectra of SN\,2020acat with those of other well observed SNe\,IIb. We then analyse the late time spectra and place constraints on the mass of oxygen synthesised by SN\,2020acat. Finally, in Section \ref{Conclusion}, we summarise the outcomes of our analysis of SN\,2020acat. 

\section{Host Galaxy and Explosion Date}
\label{Host}

\begin{figure}
    \adjustbox{trim={4.3cm, 0, 3cm, 0}, clip=true}{
    \centering
    \includegraphics[width=2\columnwidth]{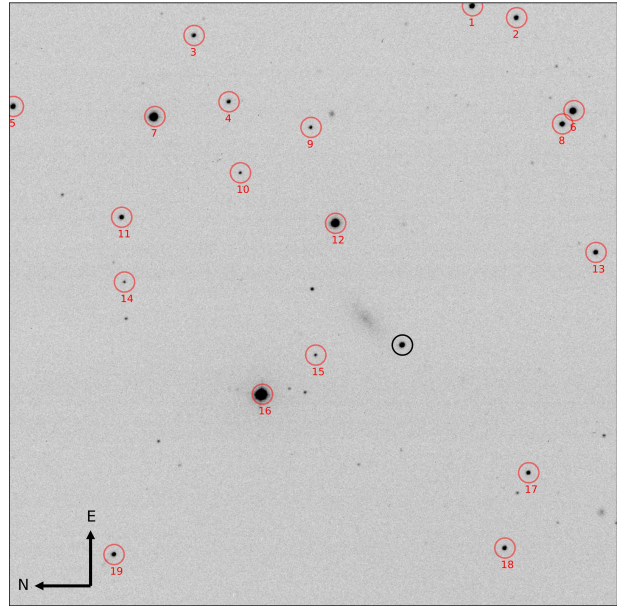}}
    \caption{$ugriz$ combined image of SN\,2020acat (black) and surrounding standard stars (red), taken on 22/12/2020 during peak light using the Liverpool Telescope. The standard stars were used to calibrate the LT $ugriz$ photometry.}
    \label{star_field}
\end{figure}

\begin{figure*}
    \centering
    \includegraphics[width=\linewidth]{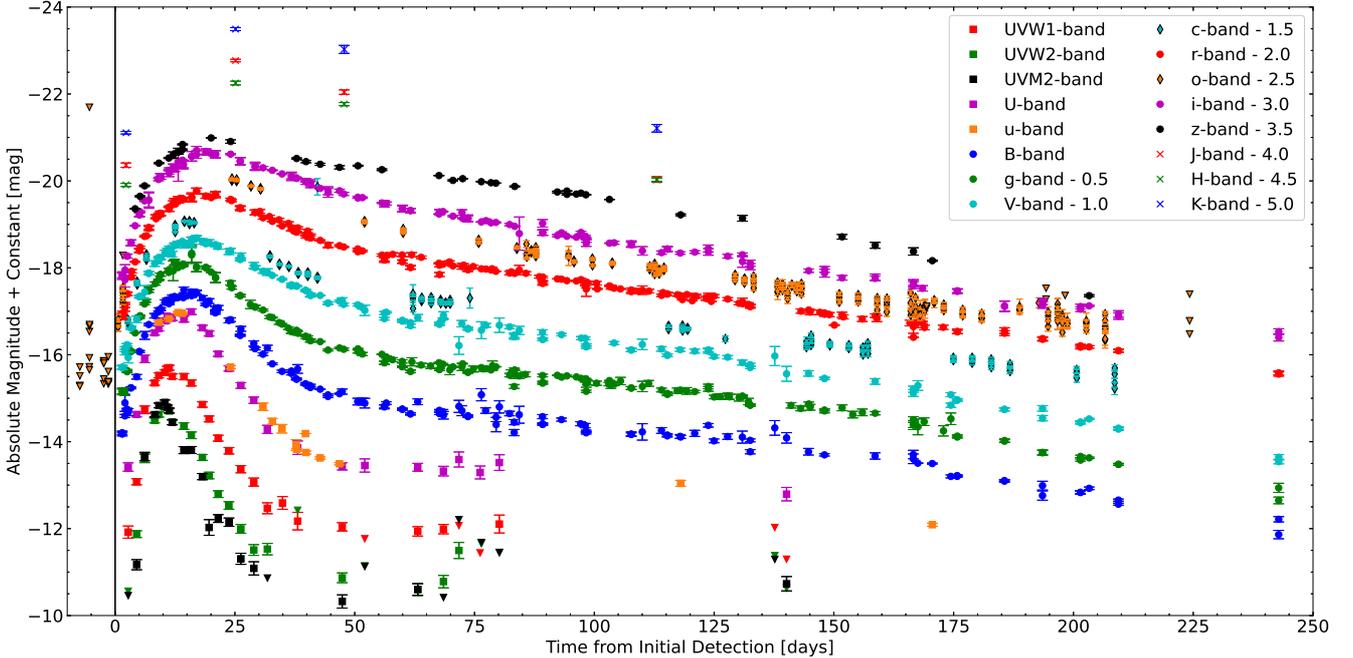}
    \caption{UV to NIR photometric observations of SN\,2020acat, with phase relative to the initial observation taken on MJD$ = 59192.66$ and given in rest frame. Each photometric band is given a marker signifying its location on the electromagnetic spectrum; UV = square, Optical = circle, and NIR = crosses. ATLAS bands are given by the coloured diamonds with the black outline and are separate from the other optical bands due to the broad nature of the ATLAS filters. The estimated explosion date is given by the solid black line.}
    \label{All_phot}
\end{figure*}

SN\,2020acat was discovered in the galaxy PGC037027 \citep{2020TNSAN.249....1S}, a $W1 = 14.50 \pm 0.03$ mag galaxy \citep{2013wise.rept....1C}, at a redshift of $z = 0.007932 \pm 0.000150$. SN\,2020acat was located $26.70"$ South and $18.00"$ West of the galactic centre, at a projected distance of $\sim \! \! 5.7$ kpc. Figure \ref{star_field} shows the location of SN\,2020acat and the surrounding stars that were used for photometric calibration. Using the NASA/IPAC Extragalactic Database (NED) default cosmology of $H_\mathrm{0} = 73.0 \pm 5\, \mathrm{kms^{-1}Mpc^{-1}} \text{, } \Omega_\mathrm{matter} = 0.27 \text{ and } \Omega_\mathrm{vacc} = 0.73$ \citep{2007ApJS..170..377S}, the host galaxy distance was found to be $35.3 \pm 2.6 \, \mathrm{Mpc}$ as derived from the model based on the local velocity given in  \citet{2000ApJ...529..786M}. The recession velocity for the host galaxy of SN\,2020acat has a velocity of $2593 \pm 53$ \kms, calculated using the same method as the distance. From this host galaxy distance, the implied distance modulus for SN\,2020acat is $m - M = 32.74 \pm 0.03$ mag. 

The line of sight dust extinction of the host galaxy, $E(B-V)_\mathrm{host}$, associated with SN\,2020acat, is expected to be negligible. No strong, narrow interstellar \NaI D lines are detected at the redshift of the host galaxy, see Section \ref{spec}. This lack of strong Na I D lines, along with the position of SN\,2020acat relative to its host galaxy, strongly suggests that $E(B-V)_\mathrm{host}$ to be negligible. 
The Milky Way (MW) extinction takes a value of $E(B-V)_\mathrm{MW} = 0.0207 \pm 0.0004$ mag, derived from the \citet{2011ApJ...737..103S} dust map. As such, we assume a total extinction for SN\,2020acat to be $E(B-V)_\mathrm{tot} = 0.0207 \pm 0.0004$ mag.

The first detection of SN\,2020acat, taken on $\mathrm{MJD} = 59192.65$ (09/12/20), occurred almost exactly two days after that last non-detection taken by the Asteroid Terrestrial-impact Last Alert System \citep[ATLAS;][]{Tonry_2018, Smith_2020} on $\mathrm{MJD} = 59190.61$ (07/12/20). This last non-detection had a limiting magnitude of $19.33 $ mag in the ATLAS $o$-band, approximately $\sim \! \! 0.78 $ mag dimmer than the initial observation. The last non-detection places a strong constraint on the explosion date and suggests that SN\,2020acat was caught very close to it. From fitting the pseudo-bolometric light curve, see Section \ref{bol_sec}, with a modified Arnett-like model, an explosion date of $\mathrm{MJD} = 59192.01 \pm 0.14$ was determined. This estimate is taken as the explosion date throughout this work. 

\section{Data Acquisition}
\label{aquis}
\subsection{Photometry}
\label{phot_aquis}

Photometry of SN\,2020acat was obtained in the UV $(UVW2, UVM2, UVW1, u, U)$, optical $(BgVriz)$ and NIR $(JHK)$ photometric bands, displayed in Figure \ref{All_phot}. SN\,2020acat was initially detected by ATLAS on $\mathrm{MJD} = 59192.66$ (09/12/20), which then continued to follow SN\,2020acat in the ATLAS $o+c$-bands for $\sim \! \! 210$ days. Optical follow-up of SN\,2020acat in the $BgVriz$-bands was obtained from several telescopes over the campaign lasting $\sim \! \! 230$ days. These telescopes include the 2.0m Liverpool Telescope \citep[LT;][]{2004SPIE.5489..679S}, the 2.56m Nordic Optical Telescope with the Alhambra Faint Object Spectrograph and Camera (ALFOSC) and the 1.82m Copernico Asiago Telescope (CT) with AFOSC. Additional photometry was provided by the Palomar 1.2m Samuel Oschin telescope using the Zwicky Transient Facility camera \citep[ZTF-cam;][]{Bellm_2018}, the 67/92 cm Schmidt telescope using Moravian and several telescopes as part of the Las Cumbres Observatory \citep[Las Cumbres;][]{2013PASP..125.1031B} through the Global Supernova Project \citep[GSP][]{2019AAS...23325816H}. All data reduction was done by automatic pipelines associated with each telescope group, with photometric magnitudes obtained through Point Spread Function (PSF) photometry. UV photometry of SN2020acat was obtained by the Neil Gehrels \textit{Swift} Observatory \citep[\textit{SWIFT};][]{Roming_2005} between 11/12/20 and 28/04/21. The UVOT data were reduced using the standard pipeline available in the HEAsoft software package \footnote{\url{https://heasarc.nasa.gov/lheasoft/}} using the latest version of CALDB. Observation of every epoch was conducted using one or several orbits. To improve the signal-to-noise ratio of the observation in a given band in a particular epoch, we have co-added all orbit-data for that corresponding epoch using the HEAsoft routine \texttt{uvotimsum}. We have used the routine \texttt{uvotdetect} to determine the correct position of the transient (which is consistent with the ground-based optical observations) and used the routine \texttt{uvotsource} to measure the apparent magnitude of SN\,2020acat by performing aperture photometry. For source extraction we have used a small aperture of radius 3.5$\arcsec$, while an aperture of radius 100$\arcsec$ have been used to determine the background. The SN is located at the outskirt of its host, implying a negligible host contribution in the NUV bands. Moreover, as a small aperture has been used to extract the flux at the SN location, considerable host contribution is also not expected in the \textit{SWIFT} optical (`U,B,V') bands. The NIR photometry was obtained by the 2.56m Nordic Optical Telescope \citep[NOT;][]{Djupvik_2010} equipped with NOTCAM through the NOT Unbiased Transient Survey 2 (NUTS2) and the 3.58m New Technology Telescope \citep[NTT;][]{1983ESOC...17..173W} through the ESO Spectroscopic Survey for Transient Objects \citep[ePESSTO+;][]{2015A&A...579A..40S} with SOFI \citep{1998Msngr..91....9M}. The NIR follow-up campaign was not as thorough as the UV and optical. All photometry have been corrected for reddening and are given in the rest frame. 

\begin{figure*}
    \includegraphics[width=\linewidth, height=22cm]{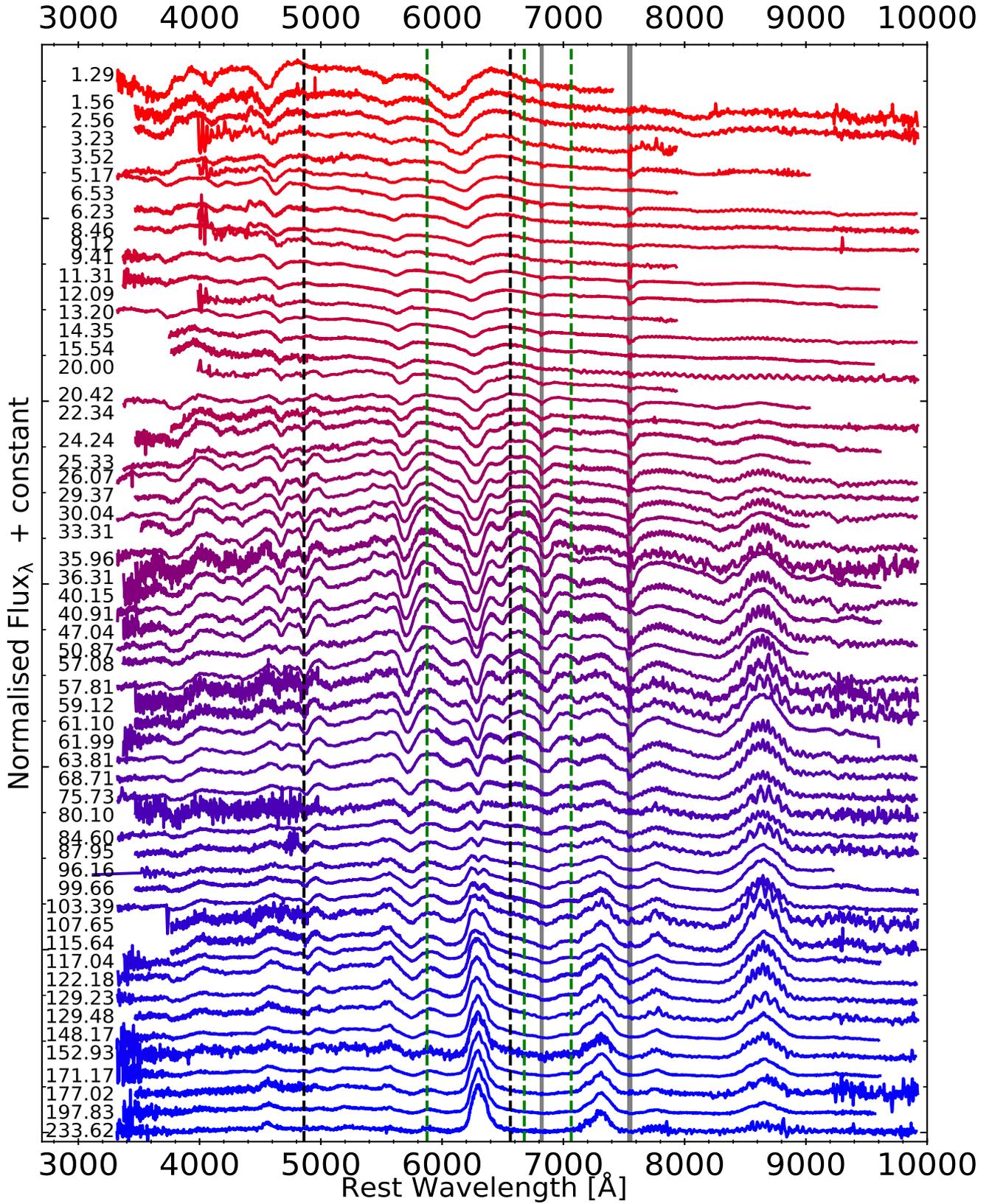} 
    \caption{Spectroscopic evolution of SN\,2020acat from multiple instruments. Flux normalised to \Ha\ feature or peak of the \OI\ $\lambda\lambda 6300,6363$ feature depending on what was stronger at the time of observation. See Table \ref{spec_info} for the details on each spectrum. The phase of each spectrum is given on the left hand side. Both the hydrogen (black) and helium (\textcolor{OliveGreen}{green}) main optical lines are given by the dashed lines, along with the telluric features which are denoted by the grey regions.}
    \label{All_spec}
\end{figure*}

\begin{table}
    \centering
    \caption{Information on spectroscopic follow-up campaign for SN\,2020acat. The epochs are relative to the explosion date ($\mathrm{MJD} = 59192.01$) and are given in the rest frame. Wavelength range is also given in rest frame. \\
    $a:$ Telescope and instrument; \\
    1. NTT = NTT using EFOSC2\\
    2. LasC = $\alpha$: FTS using FLOYDS, $\beta$: FTN using FLOYDS\\
    3. LT = LT using SPRAT \\
    4. NOT = NOT using ALFOSC\\
    5. UH88 = UH88 using SNIFS \\
    6. ASI = CT using AFOSC }
    \resizebox{\columnwidth}{!}{%
    \begin{tabular}{ccccc}
    \hline
    Spectrum & Epoch & Observation date & Telescope$^a$ & Range \\
    & $[days]$ & $[UT]$ & & [$\si{\angstrom}$] \\
    \hline
1 & 1.29 & 07:21:47 10/12/20 & NTT & 3323 - 7406  \\
2 & 1.56 & 13:56:22 10/12/20 & LasC$_{\alpha}$ & 3470 - 9921  \\
3 & 2.56 & 14:13:23 11/12/20 & LasC$_{\alpha}$ & 3470 - 9922  \\
4 & 3.23 & 06:20:02 12/12/20 & LT & 3988 - 7931  \\
5 & 3.52 & 13:23:36 12/12/20 & UH88 & 3376 - 9028  \\
6 & 5.17 & 05:20:31 14/12/20 & LT & 3988 - 7931  \\
7 & 6.23 & 06:59:30 15/12/20 & NTT & 3323 - 9910  \\
8 & 6.53 & 14:12:06 15/12/20 & LasC$_{\alpha}$ & 3471 - 9921  \\
9 & 8.46 & 12:54:15 17/12/20 & LasC$_{\alpha}$ & 3470 - 9921  \\
10 & 9.12 & 04:51:08 18/12/20 & LT & 3988 - 7931  \\
11 & 9.41 & 11:53:10 18/12/20 & NOT & 3370 - 9600  \\
12 & 11.31 & 09:51:05 20/12/20 & NOT & 3371 - 9580  \\
13 & 12.09 & 04:47:46 21/12/20 & LT & 3988 - 7931  \\
14 & 13.20 & 07:34:23 22/12/20 & NTT & 3323 - 9910  \\
15 & 14.35 & 11:25:59 23/12/20 & NOT & 3751 - 9556  \\
16 & 15.54 & 16:06:27 24/12/20 & LasC$_{\beta}$ & 3769 - 9922  \\
17 & 20.00 & 03:57:18 29/12/20 & LT & 3988 - 7931  \\
18 & 20.42 & 14:18:10 29/12/20 & UH88 & 3376 - 9028  \\
19 & 22.33 & 12:35:17 31/12/20 & LasC$_{\alpha}$ & 3767 - 9922  \\
20 & 24.24 & 10:30:45 02/01/21 & NOT & 3473 - 9612  \\
21 & 25.33 & 12:54:04 03/01/21 & UH88 & 3376 - 9028  \\
22 & 26.07 & 06:54:28 04/01/21 & NTT & 3323 - 9910  \\
23 & 29.38 & 14:46:34 07/01/21 & LasC$_{\alpha}$ & 3471 - 9921  \\
24 & 30.04 & 06:56:26 08/01/21 & NTT & 3323 - 9910  \\
25 & 33.31 & 13:54:24 11/01/21 & UH88 & 3518 - 9019  \\
26 & 35.96 & 06:05:24 14/01/21 & NTT & 3323 - 9910  \\
27 & 36.31 & 14:31:12 14/01/21 & LasC$_{\beta}$ & 3471 - 9922  \\
28 & 40.15 & 11:34:23 18/01/21 & ASI & 3374 - 9610  \\
29 & 40.91 & 05:54:07 19/01/21 & NTT & 3323 - 9910  \\
30 & 47.04 & 10:09:25 25/01/21 & NOT & 3372 - 9615  \\
31 & 50.87 & 06:53:36 29/01/21 & NTT & 3325 - 9911  \\
32 & 57.08 & 12:59:24 04/02/21 & UH88 & 3369 - 9010  \\
33 & 57.81 & 06:40:43 05/02/21 & NTT & 3325 - 9912  \\
34 & 59.12 & 14:29:24 06/02/21 & LasC$_{\beta}$ & 3471 - 9921  \\
35 & 61.10 & 14:11:41 08/02/21 & LasC$_{\beta}$ & 3469 - 9922  \\
36 & 61.99 & 11:50:35 09/02/21 & NOT & 3371 - 9601  \\
37 & 63.81 & 07:52:36 11/02/21 & NTT & 3325 - 9911  \\
38 & 68.71 & 06:16:37 16/02/21 & NTT & 3323 - 9910  \\
39 & 75.73 & 08:17:39 23/02/21 & NTT & 3327 - 9911  \\
40 & 80.09 & 17:49:21 27/02/21 & LasC$_{\beta}$ & 3470 - 9921  \\
41 & 84.60 & 06:41:28 04/03/21 & NTT & 3331 - 9911  \\
42 & 87.95 & 15:46:53 07/03/21 & LasC$_{\beta}$ & 3471 - 9920  \\
43 & 96.16 & 10:15:06 16/03/21 & ASI & 3124 - 9224  \\
44 & 99.66 & 11:03:20 19/03/21 & LasC$_{\alpha}$ & 3470 - 9921  \\
45 & 103.39 & 05:12:08 23/03/21 & NTT & 3327 - 9911  \\
46 & 107.65 & 12:14:05 27/03/21 & LasC$_{\beta}$ & 3767 - 9922  \\
47 & 115.64 & 13:37:24 04/04/21 & LasC$_{\beta}$ & 3768 - 9922  \\
48 & 117.04 & 23:23:00 05/04/21 & NOT & 3376 - 9611  \\
49 & 122.18 & 03:45:14 11/04/21 & NTT & 3327 - 9911  \\
50 & 129.23 & 06:23:16 18/04/21 & NTT & 3323 - 9910  \\
51 & 129.48 & 12:24:11 18/04/21 & LasC$_{\beta}$ & 3470 - 9922  \\
52 & 148.17 & 08:26:10 07/05/21 & NOT & 3370 - 9600  \\
53 & 152.93 & 03:40:55 12/05/21 & NTT & 3318 - 9904  \\
54 & 171.17 & 12:54:07 30/05/21 & NOT & 3373 - 9611  \\
55 & 177.03 & 10:27:00 05/06/21 & LasC$_{\beta}$ & 3471 - 9922  \\
56 & 197.83 & 09:42:37 26/06/21 & NOT & 3373 - 9569  \\
57 & 233.62 & 11:30:24 01/08/21 & NTT & 3319 - 9910  \\
    \hline
    \end{tabular}%
    }
    \label{spec_info}
\end{table}

\begin{table}
    \centering
    \caption{MJD dates and epochs relative to explosion date of maximum light in UV-NIR photometric bands for SN\,2020acat. The apparent magnitude at peak is also given for each band. All epochs are given in rest frame. The NIR bands ($JHK$) have insufficient date around peak to determine the rise time and peak epoch.}
    \resizebox{\columnwidth}{!}{%
    \begin{tabular}{ccccc}
        \hline
        band & $t_\mathrm{peak}$ & Rise time $(t_r)$ & $m_\mathrm{peak}$ & $M_\mathrm{peak}$ \\
         & $[MJD]$ & $[days]$ & $[mag]$ & $[mag]$ \\
        \hline
        $ UVW2 $ & $ 59202.20 \pm 0.64 $ & $ 10.19 \pm 0.40 $ & $ 18.20 \pm 0.03 $& $ -14.72 \pm 0.03 $ \\
        $ UVM2 $ & $ 59202.36 \pm 0.64 $ & $ 10.35 \pm 0.40 $ & $ 18.09 \pm 0.04 $& $ -14.84 \pm 0.03 $ \\
        $ UVW1 $ & $ 59203.54 \pm 0.66 $ & $ 11.53 \pm 0.42 $ & $ 17.32 \pm 0.03 $& $ -15.64 \pm 0.02 $ \\
        $ u $ & $ 59204.77 \pm 0.47 $ & $ 12.76 \pm 0.20 $ & $ 15.89 \pm 0.02 $& $ -16.95 \pm 0.02 $ \\
        $ U $ & $ 59204.88 \pm 0.66 $ & $ 12.87 \pm 0.41 $ & $ 15.83 \pm 0.02 $& $ -17.01 \pm 0.02 $ \\
        $ B $ & $ 59207.19 \pm 0.88 $ & $ 15.18 \pm 0.75 $ & $ 15.42 \pm 0.04 $& $ -17.40 \pm 0.01 $ \\
        $ g $ & $ 59208.06 \pm 0.92 $ & $ 16.05 \pm 0.82 $ & $ 15.16 \pm 0.01 $& $ -17.65 \pm 0.01 $ \\
        $ V $ & $ 59208.65 \pm 0.85 $ & $ 16.64 \pm 0.71 $ & $ 15.18 \pm 0.01 $& $ -17.62 \pm 0.01 $ \\
        $ r $ & $ 59210.20 \pm 0.88 $ & $ 18.19 \pm 0.75 $ & $ 15.09 \pm 0.01 $& $ -17.70 \pm 0.01 $ \\
        $ i $ & $ 59212.17 \pm 1.10 $ & $ 20.16 \pm 1.18 $ & $ 15.14 \pm 0.01 $& $ -17.64 \pm 0.01 $ \\
        $ z $ & $ 59213.09 \pm 1.53 $ & $ 21.08 \pm 2.29 $ & $ 15.23 \pm 0.05 $& $ -17.44 \pm 0.05 $ \\
        \hline
    \end{tabular}%
    }
    \label{phot_peaks}
\end{table}

\subsection{Spectroscopy}
\label{spect_aquis}
SN\,2020acat was classified as a Type IIb SN on $\mathrm{MJD} = 59193.31$ (10/12/2020) \citep[][]{2020TNSAN.248....1P} using a spectrum obtained by ePESSTO+ using the ESO Faint Object Spectrograph and Camera \citep[EFOSC2;][]{1984Msngr..38....9B} mounted on the NTT. The classification spectrum was obtained $\sim \! \! 1.1$ days after the explosion. Further optical spectroscopic observations of SN\,2020acat were obtained from ePESSTO+ using the EFOSC2 camera with the NTT. Spectroscopic observations from ePESSTO+ were obtained with the blue grism, gr11 $(3380 - 7520 \, \si\angstrom)$, and the red grism, gr16 $(6015 - 10320 \, \si\angstrom)$. These EFOSC2 spectra were then combined to form a single spectrum at each epoch with full optical coverage. Several additional spectra were obtained during the evolution of SN\,2020acat using the Spectrograph for the Rapid Acquisition of Transients \citep[SPRAT;][]{10.1117/12.2055117} mounted on the LT, the Supernova Integral Field Spectrograph \citep[SNIFS;][]{2004SPIE.5249..146L} camera mounted on the Hawaii based UH88 telescope and via the NUTS2 programme using the ALFOSC mounted on NOT using grism 4. Additional spectra were obtained through Las Cumbres using the FLOYDS spectrographs mounted on the 2\,m Faulkes Telescope South (FTS) and the 2\,m Faulkes Telescope North (FTN), based at the Siding Spring Observatory (COJ) and the Haleakala Observatory (OGG), respectively. Spectra was also obtained using the 1.82m Copernico telescope using AFOSC, with both the VPH7 and VPH6 grisms. All spectra were reduced in the standard procedure for each telescope. The spectroscopic follow-up campaign that lasted for $\sim \! \! 230$ days before SN\,2020acat was no longer observable. The spectroscopic evolution of SN\,2020acat can be seen in Figure \ref{All_spec}, with details on the spectroscopic observations given in Table \ref{spec_info}.

\section{Photometry Analysis}
\label{phot}

\begin{figure*}
    \centering
    \includegraphics[width=\linewidth]{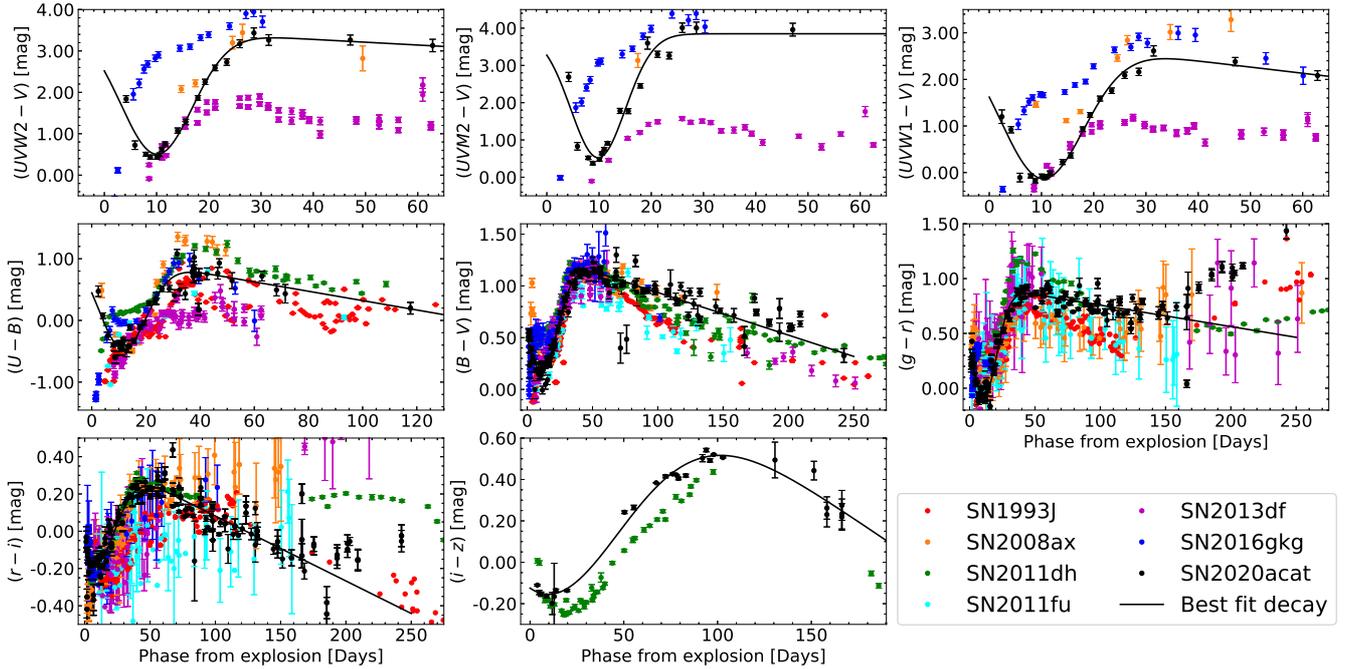}
    \caption{The $(UVW2-V)$, $(UVM2-V)$, $(UVW1-V)$, $(U-B)$, $(B-V)$, $(g-r)$, $(r-i)$ and $(i-z)$ colour evolution of SN\,2020acat (black) compared to the colour evolution of SN\,1993J (\textcolor{red}{red}), SN\,2008ax (\textcolor{orange}{orange}), SN\,2011dh (\textcolor{OliveGreen}{green}), SN\,2011fu (\textcolor{cyan}{cyan}), SN\,2013df (\textcolor{Fuchsia}{magenta}), SN\,2016gkg (\textcolor{blue}{blue}). The colour evolution of SN\,2020acat was fit with a decay function (black line) which assumes a linear decline at late times $\gtrsim 35$ days. All SNe colours are given in the rest frame.}
    \label{Colour_evo}
\end{figure*}

\subsection{UV to NIR Light Curves}
The UV bands of SN\,2020acat were only followed for $\sim \! \! 135$ days before becoming too dim to observe. The optical bands were followed for a total of $\sim \! \! 250$ days, with the NIR bands being observed a few times throughout the follow-up campaign. The rise of the UV and optical light curves was observed with a fast cadence. The proximity of the initial observation to the last non-detection for SN\,2020acat, along with the depth of the limit in the ATLAS $o$-band, argues against the possibility of a bright long duration shock cooling phase occurring such as those seen in SN\,1993J \citep{10.1093/mnras/266.1.L27}, SN\,2011dh \citep{2011ApJ...742L..18A} or SN\,2016gkg \citep{2017ApJ...837L...2A}. The lack of a long duration shock cooling tail was also confirmed by the UV bands, which tentatively show the cooling tail if it is present, with the first UV observation taken $\sim \! \! 2.5$ days after the estimated explosion no deviation from the fast rise is seen in all the other bands. 

The peaks of all but the NIR bands were well observed, allowing constraints to be placed on the epoch of maximum brightness and the value for peak magnitude for each band. The epoch of maximum brightness in each band, along with the MJD date, peak apparent and absolute magnitude are given in Table \ref{phot_peaks}. The epochs of maximum light were determined by fitting a cubic spline to the light curves of each photometry band around peak time. The error associated with the peak time is a combination of the error from the explosion date and the fitting of the cube spline. Unfortunately, the reduced number of observations in $z$-band around maximum light resulted in the epoch of peak brightness to have an error $\sim \! \! 2$ times greater than the other bands. It should be noted that, due to the lack of any $JHK$-band data around peak time, a spline could not be fitted without placing an extremely large uncertainty to the epochs of maximum light.

SN\,2020acat has one of the fastest rise times among SNe\,IIb, with the UV bands peaking in $\sim \! \! 10$ days, while the optical bands peaking in $\sim \! \! 14 - 22$ days. This is faster than the average SNe IIb which reaches peak in the UV bands in $\gtrsim \! \! 15$ days, with the optical bands reaching peak in $\gtrsim \! \! 20$ days, as seen with SN\,2008ax \citep{Roming_2009} and SN\,2011dh \citep{Marion_2014}. The UV bands also display a very fast decline once they reach peak light with an average value for $\Delta m^{UV}_{15} = 2.35 \pm 0.04$ mag, while the optical bands decline with an average $\Delta m^{opt}_{15} = 0.77 \pm 0.24$ mag, suggesting that the ejecta of SN\,2020acat rapidly expanded and cooled scattering the light to lower energy.
The $B$-band of SN\,2020acat had a rise time of $B_{t_r}= 15.18 \pm 0.75$ days, roughly $\sim \! \! 4$ days faster than for SN\,1993J \citep[$B_{t_r} = 18.97$ days, ][]{1994AJ....107.1022R}, SN\,2008ax \citep[$B_{t_r} = 18.9$ days, ][]{2008MNRAS.389..955P}, SN\,2011dh \citep[$B_{t_r} = 19.6$ days, ][]{2013MNRAS.433....2S} and SN\,2011fu \citep[$B_{t_r} = 23.23$ days, ][]{10.1093/mnras/stt162}. The short photometric rise time, seen in all bands of SN\,2020acat, is characteristic of SNe Type II with small progenitor radii \citep{10.1093/mnras/stv1097}, suggesting that SN\,2020acat might have originated from a compact progenitor. The compact progenitor idea is further enhanced by the lack of any visible shock-breakout phase, characteristic of more compact objects. 

\subsection{Colour Evolution}

The $(UVW2-V)$, $(UVM2-V)$, $(UVW1-V)$, $(U-B)$, $(B-V)$, $(g-r)$ $(r-i)$ and $(i-z)$ colours evolution of SN\,2020acat are given in Figure \ref{Colour_evo}. For the $(U-B)$ colour evolution, additional data was added using the $(u-g)$ and $(B-V)$ colours which forms a relation with $(U-B)$, as given by \citet{SDSS_conversion}, of;
\begin{ceqn}
\begin{align}
    (U-B) = \frac{(u-g) - (0.770\pm0.05)(B-V)-(0.72\pm0.04)}{(0.75\pm0.05)}.
    \label{U-B}
\end{align}
\end{ceqn}

\begin{figure*}
    \centering
    \includegraphics[width=\linewidth]{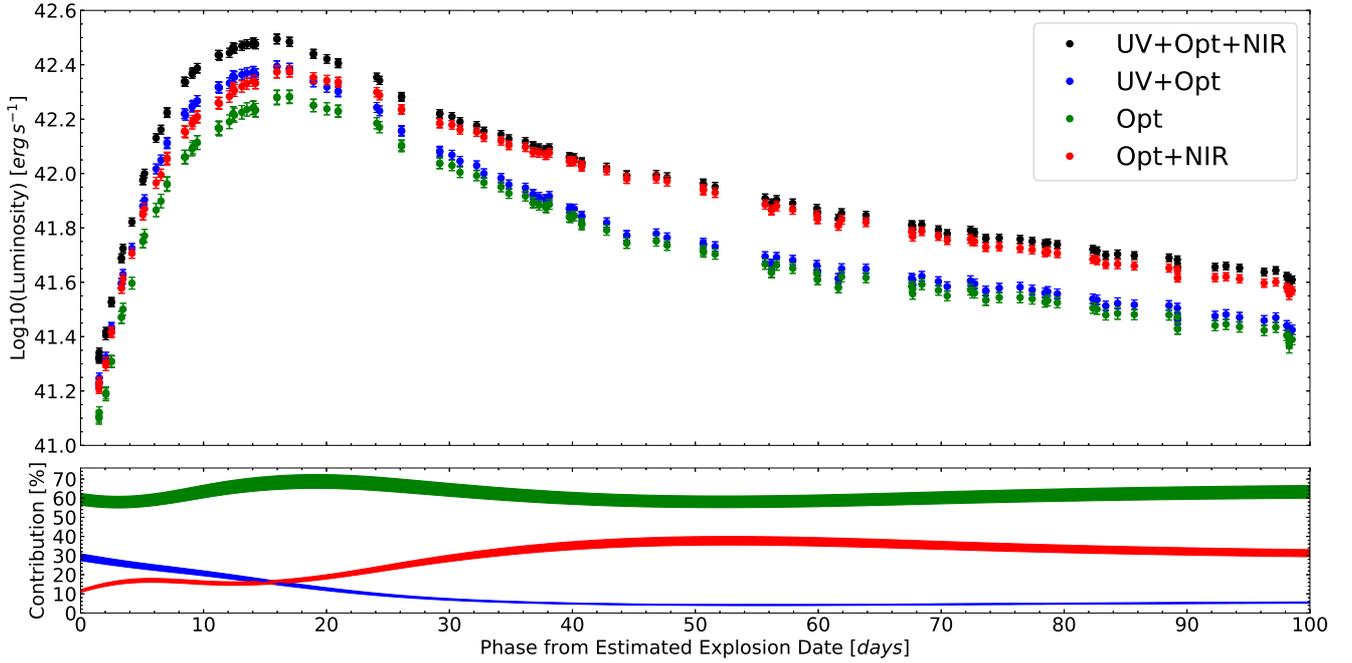}
    \caption{(Top) The Pseudo-bolometric light curve of SN\,2020acat, along with the light curves constructed using the UV+optical, just optical and optical+NIR photometry. (Bottom) The contribution as a percentage of the pesudo-bolometric light curve of the individual electromagnetic regions, with UV = \textcolor{blue}{blue}, Optical = \textcolor{OliveGreen}{green} and NIR = \textcolor{red}{red}.}
    \label{Bol+cont}
\end{figure*}

The colours obtained from the UV and optical photometry of SN\,2020acat were all fit with a combination of a Gaussian function, to follow the initial decline and rise seen within all colours during the first few weeks, and a linear decay function, which fits to the late time colour decline. The decay function is shown as a solid black line in Figure \ref{Colour_evo} and was only fit to the colour evolution of SN\,2020acat. Initially the colours of SN\,2020acat start relatively red before rapidly decreasing towards bluer colours within the first two weeks, reaching minima at around day $\sim \! \! 10-15$. After this point, the colours evolve redwards as they rise to peak at $\sim \! \! 35 - 45$ days. Once this peak is reached, the colour in all bands slowly declines over the next $\sim \! \! 135$ days. The $(U - B)$ colour declines at a much faster rate, $\sim \! \! 0.007 \text{mag/day}$, than the other colours, with the $(B - V)$, $(g - r)$ and $(r - i)$ having a decline of $\sim \! \! 0.004$, $\sim \! \! 0.002$ and $\sim \! \! 0.003 $ mag per day, respectively. The fast decline of the $(U - B)$ colour results from the rapid dimming that the UV bands undergo within the first few weeks of maximum light as the ejecta expands and cools scattering the UV photons towards the redder photometric bands, a feature seen for all SNe\,IIb shown in Figure \ref{Colour_evo}. While both the $(U - B)$ and $(B - V)$ colours follow a linear decline at the late time ($t > 150$ days), both the $(g - r)$ and $(r - i)$ colours diverge from a purely linear decline. The $(g - r)$ colour starts to rapidly grow redder again, while the $(r - i)$ colour evolution seems to levels out, as SN\,2020acat cools and the light curve becomes dominated by the redder bands. More observations at later times would have been needed to see the true evolution of the $(r - i)$ colour, as the deviation from a linear decline is not as strong as for the $(g - r)$ colour.

The multi-band colour evolution of SN\,2020acat is compared to those of SN\,1993J \citep{1994AJ....107.1022R, 1995A&AS..110..513B, 1996AJ....112..732R}, SN\,2008ax \citep{2008MNRAS.389..955P, 2009PZ.....29....2T, 2011MNRAS.413.2140T}, SN\,2011dh \citep{tsvetkov2012photometric, 2013MNRAS.433....2S, 2014Ap&SS.354...89B, 2014A&A...562A..17E} and SN\,2016gkg \citep{2014Ap&SS.354...89B, 2017ApJ...837L...2A, 2018Natur.554..497B}. These SNe\,IIb were chosen as comparison objects for SN\,2020acat as they all possess comprehensive photometric and spectroscopic data around peak time, as well as into the late time when the hydrogen features have faded. The thoroughly documented nature of these events mean they have well known properties, allowing a comprehensive comparison between the results obtained from the Arnett-like model described below and the literature values. This is used as a test to validate the model used here and thus the results obtained for SN\,2020acat. The colour evolutions of these SNe\,IIb are also given in Figure \ref{Colour_evo}, with details of each SN\,IIb given in Table \ref{SNe_IIb info}. For several SNe only Johnson-Cousins photometric bands were available for the redder ($r$ to $i$) bands, such as SN\,1993J. For these SNe a conversion to Sloan Digital Sky Survey (SDSS) red bands was done using equation \ref{g-r} and \ref{r-i} also from \citet{SDSS_conversion},

\begin{ceqn}
\begin{align}
    (g-r) = (1.646\pm0.008)(V-R) - (0.139\pm0.004),
    \label{g-r}
\end{align}
\end{ceqn}

\noindent and 

\begin{ceqn}
\begin{align}
    (r-i) = (1.007\pm0.005)(R-I) - (0.236\pm0.003).
    \label{r-i}
\end{align}
\end{ceqn}

\begin{figure*}
    \centering
    \includegraphics[width=\linewidth]{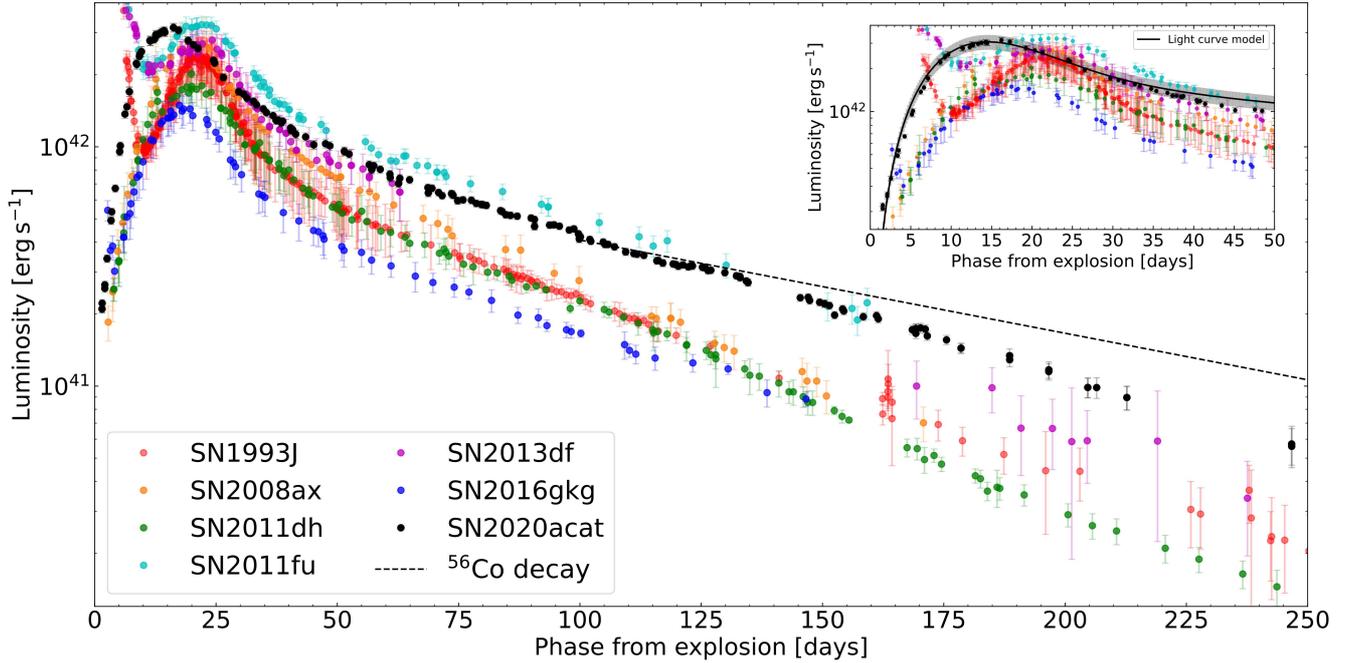}
    \caption{The pseudo-bolometric light curve of SN\,2020acat, along with several other well observed SNe\,IIb over the 250 day period from explosion. The SNe shown in the plot are SN\,1993J (\textcolor{red}{red}), SN\,2008ax (\textcolor{orange}{orange}), SN\,2011dh (\textcolor{OliveGreen}{green}), SN\,2011fu (\textcolor{cyan}{cyan}), SN\,2013df (\textcolor{Fuchsia}{magenta}), SN\,2016gkg (\textcolor{blue}{blue}) and SN\,2020acat (black). All SNe have been corrected for reddening as well as time dilation. The black dashed line displays the cobalt decay line that should dominate at late times for a light curve powered solely by the decay of \Nifs. The subplot (upper right) displays the initial 50 days of each pseudo-bolometric light curves. For SN\,2020acat the Arnett-like function is also displayed by the solid black line, along with associated errors for the model shown by the grey shaded region. The physical parameters of each SN\,IIb from the Arnett-like fit are given in Table \ref{Physical_params}.}
    \label{Bol_comp}
\end{figure*}

While a comprehensive comparison of the colour evolution's of SN\,2020acat can be done with the majority of the optical bands, the same cannot be said for the UV colour's and the $(i-z)$ evolution. This is due to the lack of UV and $z$-band data for several of these SN. Hence, a thorough comparison with SN\,2020acat cannot be done for these bands. 
The colour evolution of SN\,2020acat initially shares the same shape as SN\,2008ax, displaying an initial decline to a bluer colour before rising to a red peak within $\sim \! \! 4$ weeks of the explosion, as expected for a SN that lacks detection of a strong shock-cooling tail, and thus displays an initial very blue colour before becoming redder. After the red maximum at $\sim \! \! 4$ weeks post-explosion, the colour evolution of SN\,2020acat follows the shape of SN\,1993J and SN\,2011dh, which both display a decline for several months before changing slope, although the rate of decline of SN\,2020acat is much slower than the other SNe. At around $\sim \! \! 140 - 150$ days after explosion several SNe, including SN\,2020acat, diverge from a linear decay in the $(g-r)$ and $(r-i)$ bands. The $(g-r)$ colour of SN\,2008ax also displays this increase, although the rise is at a much slower pace compared to both SN\,1993J and SN\,2020acat. The $(r-i)$ colour evolution of SN\,2020acat is quite similar to both SN\,1993J and SN2011fu in the decline phase. Interestingly, the shape of the $(i-z)$ colour evolution of SN\,2020acat is very similar to that of SN\,2011dh, despite the clear differences seen in the bluer colours. The mentioned colour evolution of the respective comparison SNe show an initial decline followed by a rise and final decline at $> \! \! 100$ days.
When only considering the $(i-z)$ colour, it is shown that SN\,2020acat reaches a blue peak $\sim \! \! 10$ days before SN\,2011dh and is also redder throughout the entire evolution.

\subsection{Bolometric light curve}
\label{bol_sec}

\begin{table}
    \centering
    \caption{Details for the SNe\,IIb that are compared with SN\,2020acat. \\
    \scriptsize Sources: 1. \citet{1994AJ....107.1022R}, 2. \citet{1995A&AS..110..513B}, 3. \citet{1996AJ....112..732R}, 4. \citet{2008MNRAS.389..955P}, 5. \citet{2009PZ.....29....2T} 6. \citet{tsvetkov2012photometric}, 7. \citet{2013MNRAS.433....2S}, 8. \citet{10.1093/mnras/stt162}, 9. \citet{2014Ap&SS.354...89B}, 10. \citet{Morales_Garoffolo_2014}, 11. \citet{Van_Dyk_2014}, 12. \citet{2017ApJ...837L...2A}, 13. \citet{2018Natur.554..497B}. }
    \resizebox{\columnwidth}{!}{%
    \begin{tabular}{ccccccc}
        \hline
        SN & Explosion & Redshift & Distance & $E(B-V)_\mathrm{MW}$ & $E(B-V)_\mathrm{Host}$ & Source \\
         & date [MJD] & & [Mpc] & [mag] & [mag] & \\
        \hline
        1993J & 49072.0 & -0.00113 & 2.9 & 0.069 & 0.11 & 1,2,3 \\
        2008ax & 54528.8 & 0.00456 & 20.25 & 0.022 & 0.278 & 4,5 \\
        2011dh & 55712.5 & 0.001638 & 7.80 & 0.035 & 0.05 & 6,7,9 \\
        2011fu & 55824.5 & 0.001845 & 74.5 & 0.068 & 0.035 & 8 \\
        2013df & 56447.8 & 0.00239 & 21.4 & 0.017 & 0.081 & 10, 11 \\ 
        2016gkg & 57651.2 & 0.0049 & 21.8 & 0.0166 & 0.09 & 12,13 \\
        2020acat & 59192.01 & 0.007932 & 35.32 & 0.0207 & - & - \\
        \hline
    \end{tabular}%
    }
    \label{SNe_IIb info}
\end{table}

\begin{table*}
    \centering
    \caption{Physical parameters derived from fiting the pseudo-bolometric light curves for SN\,2020acat and the other SNe\,IIb with the Arnett-like fit, along with details derived from the light curve. Also shown are the literature values for the \mni, \mej and \ek of the different SNe\,IIb, these values were derived using the Arnett-like model and an optical opacity $\kappa = 0.06 \, \mathrm{cm^2 g^{-1}}$.}
    \begin{tabular}[width=\linewidth]{ccccccc|ccc}
        \hline
        & & & & \multicolumn{3}{c|}{This work} & \multicolumn{3}{c}{Literature} \\
        \multirow{2}{*}{SN} & $Log(L_\mathrm{peak})$ & $t_\mathrm{peak}$ & $v_\mathrm{ph}$ & \mni & \mej\ & \ek & \mni & \mej\ & \ek\\
        & [ \ergs ] & [$days$] & [$\time 10^{3}$\kms] & [\msun] & [\msun] & [$\times 10^{51}$\erg] & [\msun] & [\msun] & [$\times 10^{51}$\erg] \\
        \hline
        2020acat & $ 42.49 \pm 0.15 $ & $ 14.62 \pm 0.27 $ & $ 10.0 \pm 0.5 $ & $ 0.13 \pm 0.02 $ & $ 2.3 \pm 0.3 $ & $ 1.2 \pm 0.2 $ & - & - & - \\
        1993J & $ 42.37 \pm 0.17 $ & $ 19.95 \pm 0.32 $ & $ 8.0 \pm 1.0 $ & $ 0.10 \pm 0.03 $ & $ 1.9 \pm 0.4 $ & $ 0.7 \pm 0.2 $ & $ 0.10 \pm 0.04 $ & $ 2.7 \pm 0.8 $ & $ 1.3 \pm 0.3 $ \\
        2008ax & $ 42.38 \pm 0.08 $ & $ 17.20 \pm 0.27 $ & $ 7.5 \pm 0.5 $ & $ 0.13 \pm 0.04 $ & $ 2.5 \pm 1.0 $ & $ 0.8 \pm 0.3 $ & $ 0.10 \pm 0.02 $ & $ 2.7 \pm 0.5 $ & $ 1.2 \pm 0.5 $ \\
        2011dh & $ 42.15 \pm 0.11 $ & $ 20.21 \pm 0.54 $ & $ 6.5 \pm 1.0 $ & $ 0.05 \pm 0.01 $ & $ 2.2 \pm 0.4 $ & $ 0.6 \pm 0.1 $ & $ 0.07 \pm 0.01 $ & $ 2.1 \pm 0.5 $ & $ 0.8 \pm 0.2 $ \\
        2011fu & $ 42.49 \pm 0.14 $ & $ 21.15 \pm 0.73 $ & $ 8.0 \pm 1.0 $ & $ 0.17 \pm 0.03 $ & $ 3.4 \pm 0.7 $ & $ 1.3 \pm 0.3 $ & $ 0.15 $ & $ 3.5 $ & $ 1.3 $ \\
        2013df & $ 42.38 \pm 0.07 $ & $ 17.84 \pm 0.30 $ & $ 8.0 \pm 1.0 $ & $ 0.11 \pm 0.02 $ & $ 1.5 \pm 0.3 $ & $ 0.6 \pm 0.1 $ & $ 0.11 \pm 0.02 $ & $ 0.11 \pm 0.30 $ & $ 0.8 \pm 0.4 $ \\
        2016gkg & $ 42.13 \pm 0.02 $ & $ 18.47 \pm 0.13 $ & $ 8.0 \pm 1.0 $ & $ 0.06 \pm 0.01 $ & $ 1.6 \pm 0.3 $ & $ 0.6 \pm 0.1 $ & - & - & - \\
        \hline

    \end{tabular}
    \label{Physical_params}
\end{table*}

A pseudo-bolometric light curve was constructed using the UV - NIR photometric bands in order to obtain the physical parameters of SN\,2020acat. The pseudo-bolometric light curve was constructed by integrating the flux of the UV - NIR bands. During the epochs when the NIR bands are missing the magnitude is obtained by interpolating the points using a polynomial fit. Black body corrections were calculated and applied using the spectral energy distributions created from the interpolated photometric bands. Additional pseudo-bolometric light curves were constructed using the UV - optical, solely optical and optical - NIR photometry. The resulting pseudo-bolometric light curves are presented in the top panel of Figure \ref{Bol+cont}, the contribution of each photometric region to the complete pseudo-bolometric light curve given can be seen in the main panel of Figure \ref{Bol+cont}. The optical bands contribute the most to the total light curve throughout its evolution. Initially, the UV bands dominate over the NIR bands, but rapidly decline in strength as the NIR contribution increases, peaking at around $\sim \! \! 35\%$ and staying relatively constant until $\gtrsim \! \! 150$ days, when the NIR contribution rises. The contribution from the NIR photometry bands shown in Figure \ref{Bol+cont} is initially slightly inflated due to the interpolation at early times resulting in an overestimation in the strength of the NIR bands. This lack of comprehensive photometric coverage for the NIR region resulted in an increase in the final error for the pseudo-bolometric light curve. However, the lack of NIR coverage at peak time is not expected to have a significant effect on the physical parameters derived from the pseudo-bolometric light curve due to the domination of the UV and optical bands at these epochs. The error of the pseudo-bolometric light curve was also influenced by the lack of UV bands at late times ($\gtrsim \! \! 60$ days), although at this epoch the UV bands contribute little to the bolometric light curve ($\lesssim \! \! 10\%$ of total flux), and therefore is not expected to impose significant errors. The late time pseudo-bolometric light curve of SN\,2020acat is likely higher in luminosity than it should be due to an overestimation of the NIR bands at the late time when they contribute the most to the bolometric light curve.

The evolution of the pseudo-bolometric light curve of SN\,2020acat, along with the light curves of of SN\,1993J, SN\,2008ax, SN\,2011dh, SN\,2011fu, SN\,2013df and SN\,2016gkg, using all available photometric bands, are displayed in Figure \ref{Bol_comp}, with the peak time evolution of each pseudo-bolometric light curve shown in the upper right. These SNe were used as comparison objects for SN\,2020acat due to their comprehensive photometric coverage, which extends from the early time around their estimated explosion date to well in the nebular late time phase. This allows for a comparison of their pseudo-bolometric light curves with that of SN\,2020acat during both the pre-maximum and post maximum phases, as well as at late time when \Cofs\ decay dominates the light curve. All SNe shown in Figure \ref{Bol_comp} have also been thoroughly modelled and have well determined physical factors such as their distance modulus, extinction and explosion date. This permits a comparison and allows SN\,2020acat to be placed within the property distribution of SNe\,IIb. Also shown in Figure \ref{Bol_comp} is the decay slope of \Cofs, the source of power expected to dominate the late-time evolution of SNe. As expected, the decay of SNe\,IIb display a small spread at late time with all light curves significantly diverging from the slope of \Cofs\ decay \citep{10.1093/mnras/stv650}, as expected in the absence of the full trapping of gamma-rays released by the decay of \Cofs. 

The pseudo-bolometric light curve of SN\,2020acat reaches a peak luminosity of  $L_\mathrm{peak} = 3.09 _{-0.9}^{+1.28} \times 10^{42}$ \ergs, $Log(L_\mathrm{peak}) = 42.49 \pm 0.15 $ [\ergs], with a rise time of $\sim \! \! 14.6 \pm 0.3$ days. The rise time of SN\,2020acat is considerably faster compared to the other SNe\,IIb, which tend to have a rise time of $\sim \! \! 20$ days or longer, and was expected from the rapid rise seen in the UV and optical bands. The peak luminosity of SN\,2020acat was one of the higher among the SNe\,IIb compared in Figure \ref{Bol_comp}. Among the SNe of our sample, only SN\,2011fu is of similar brightness to SN\,2020acat, with a peak luminosity of $Log(L_\mathrm{peak}) = 42.49 \pm 0.14 \mathrm{[erg \, s^{-1}]}$, while SN\,2008ax has a similar luminosity to SN\,2020acat, peaking at a luminosity of $Log(L_\mathrm{peak}) = 42.38 \pm 0.57 \mathrm{[erg \, s^{-1}]}$. When compared to the mean value of peak luminosity for SNe IIb given by \citet{10.1093/mnras/stw299}, $Log(L_\mathrm{peak}) = 42.36 \pm ^{0.26} _{0.11} \mathrm{[erg \, s^{-1}]}$, SN\,2020acat is $\sim \! \! 0.12 \mathrm{[erg \, s^{-1}]}$ brighter, suggesting that a larger than average amount of \Nifs\ was synthesised during the explosion. The analysis done in \cite{10.1093/mnras/stw299} used the same cosmology as that given in Section \ref{Host} allowing for a clear comparison with SN\,2020acat to be made.

An Arnett-like model \citep{1982ApJ...253..785A} was fit to the pseudo-bolometric light curves of SN\,2020acat and the other SE-SNe to determine the mass of \Nifs\ synthesised, as well as the other physical parameters, such as the mass of material ejected (\mej) and the kinetic energy of the SNe (\ek). For the fitting of the Arnett-like model, an opacity of $\kappa = 0.06 \, \mathrm{cm^2 \, g^{-1}}$ and a dimensionless form factor from integration $\beta$ of $13.8$, a constant derived by \citet{1982ApJ...253..785A}, was used for all SNe. An opacity of $0.06 \, \mathrm{cm^2 \, g^{-1}}$ was used here as it has been established that a small optical opacity is needed for the modelling of the bolometric light curves for hydrogen rich SNe, and has been used in studies of Type IIb SNe \citep{10.1093/mnras/stv2983}. The degeneracy between the ejecta mass and the kinetic energy was broken for each SN by using the photospheric velocity, determined by the average velocity of the \FeII\ lines at around peak light. For SN\,2020acat, the photospheric velocity had a value of $\sim \! \! 10000 \pm 500$ \kms, which was determined from the absorption minimum of the \FeII\ $\lambda 5169$ line. The model used for SN\,2020acat and the other SNe\,IIb was slightly modified to determine the estimated explosion date by fitting to the pre-peak photometric data. This was modified for those SNe that displayed a shock cooling phase and produced values all within the error range given for the explosion data within the literature.

\begin{figure*}
    \centering
    \includegraphics[width=\linewidth]{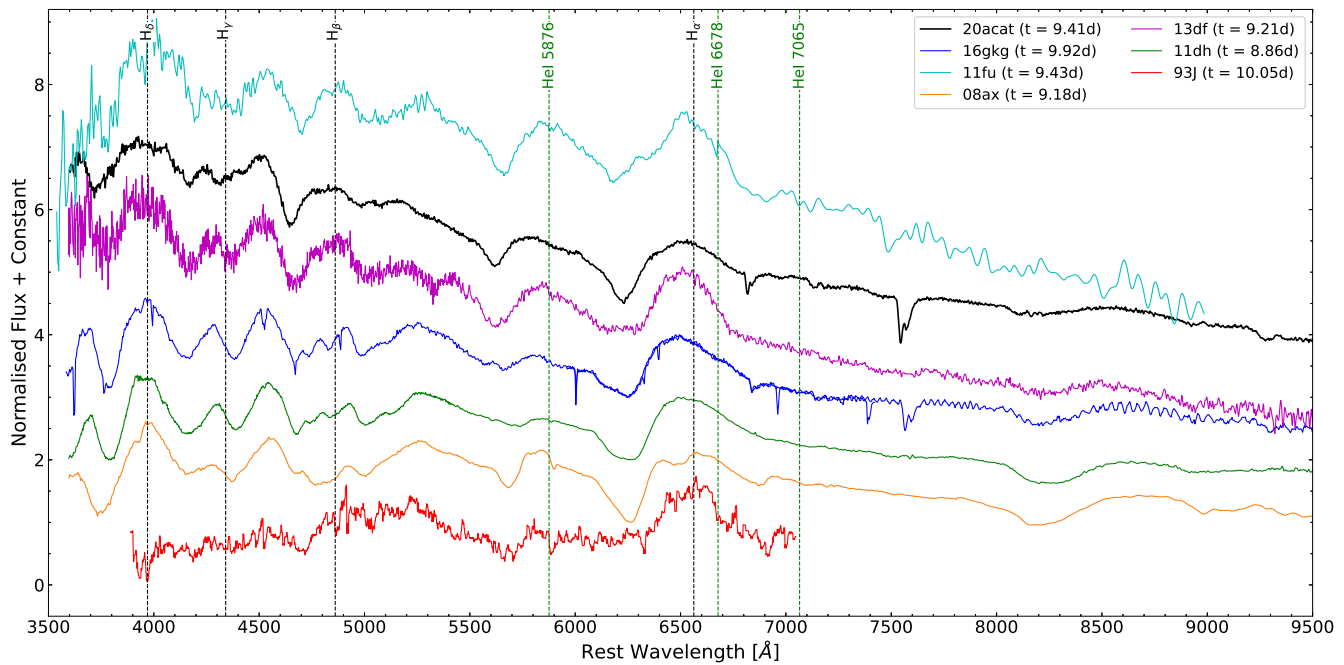}
    \caption{Spectroscopic comparison of SN\,IIb at around 9 days from explosion. All spectra have been corrected for redshift and extinction. The emerging hydrogen and helium features are marked at their rest wavelength.}
    \label{Spec_9}
\end{figure*}

The model, shown for SN\,2020acat in the upper right plot of Figure \ref{Bol_comp}, fits well to the peak of the pseudo-bolometric light curve and the majority of the rise time. The model seems to underestimate the luminosity of the initial points, within less than 5 days from explosion, and starts to diverge away from the constructed pseudo-bolometric light curve at around $35-45$ days after estimated explosion. The divergence at later times, $\geq 35$ days, is expected as the light curve transitions into being dominated by other processes and not just the decay of \Nifs. Interestingly the underestimation of the bolometric light curve at the earliest phase seems to imply that the mass of \Nifs\ is more mixed into the outer ejecta than what the model assumes. The main errors in the model arise from the error in the photospheric velocity, the error associated with the pseudo-bolometric light curve and the error within the estimated explosion data. From the fitting to the pseudo-bolometric light curve, the values for the physical parameters of SN\,2020acat and the comparison SNe\,IIb, along with the peak luminosity and rise time were determined. These values along with the photospheric velocity used to break the degeneracy between \mej\ and \ek, are given in Table \ref{Physical_params}. From the Arnett-like model fit to the light curve of SN\,2020acat, a \Nifs\ mass of \mni $= 0.13 \pm 0.02$ \msun\ was obtained, along with a ejecta mass of \mej $= 2.3 \pm 0.3$ \msun\ and a kinetic energy of \ek $= 1.2 \pm 0.2 \times 10^{51}$ \erg. 

It should be noted that in recent years there has been a lot of discussion in the validity of using an Arnett-like approach to obtain the value of the \Nifs\ mass synthesised by CC-SNe. \citet{Khatami_2019} discussed the effect of neglecting the time-dependent diffusion on the \Nifs\ mass that Arnett-like models assume. Alternative models for CC-SNe have shown that the value of \Nifs\ mass derived from the Arnett-like model is likely higher by $\sim \! \! 30 - 40 \%$ higher than the true value \citep[see,][]{10.1093/mnras/stw418, 2020arXiv200906868W}. As such if the \Nifs\ mass derived above is overestimated by $\sim \! \! 30 - 40 \%$ then the final \Nifs\ mass for SN\,2020acat would be \mni $= 0.08-0.09 \pm 0.02$ \msun. However due to the prolific use of the Arnett-like model in the literature, when comparing the physical parameters of SN\,2020acat with those of other SNe\,IIb, the uncorrected value of \mni\ will be used to give a more valid comparison.

The physical parameters obtained for SN\,2020acat suggests that it was a high energy event, producing a large amount of both \Nifs\ and ejecta. However, to get a comprehensive look at the physical parameters of SN\,2020acat a comparison with a large study of SNe\,IIb is needed. From \citet{2019MNRAS.485.1559P}, the mean \mni\ and \mej\ of SNe\,IIb were determined to be $0.07 \pm 0.03$ \msun, $2.7 \pm 1.0$ \msun\ respectively. These physical parameters show that SN\,2020acat displays a roughly average value for the ejecta mass while having a significantly higher value for the nickel mass, which would account for the brighter pseudo-bolometric light curve shown in Figure \ref{Bol_comp}. However, when compared to the analysis of SE-SNe done by \citet{10.1093/mnras/stv2983}, with values of \mni\ $= 0.11 \pm 0.04$ \msun, \mej\ $= 2.2 \pm 0.8$ \msun\ and a kinetic energy of $1.0 \pm 0.6 \times 10^{51}$ \erg, SN\,2020acat synthesised slightly more nickel than the average SNe IIb, an average amount of ejecta mass and a slightly higher kinetic energy. From both comparisons with average SNe\,IIb parameters can be seen that SN\,2020acat is an energetic event that produces a roughly average amount of ejecta for a SN\,IIb. The high energy derived for SN\,2020acat, along with the value of \mej, supports the idea that the progenitor of SN\,2020acat was an intermediate mass star with a \mzams\ between $18 - 22$ \msun. The value of the \mzams\ predicted for SN\,2020acat, while on the higher end of progenitor masses, is not out of the range of possibility for a SE-SNe. \citet{Mazzali_2003} reported a similar progenitor \mzams\, $\sim \! \! 20 - 25$ \msun, for SN\,2002ap a Type Ic SNe that produced a similar amount of ejecta mass as SN\,2020acat. A high mass progenitor, \mzams\ $\sim \! \! 18$ \msun, was also suggested by \citet{2015ApJ...811..147F} for SN\,2008ax. It was also shown by \citet{10.1093/mnras/stv2983} that the observed distribution of ejecta masses for SE-SNe can be explained by progenitors with masses that range between $8 - 20$ \msun. As such an intermediate - high mass progenitor for SN\,2020acat is not impossible, although it would require detailed hydrodynamic modelling to determine its validity, which is beyond the scope of this work.

\section{Spectroscopic Analysis}
\label{spec}
Figure \ref{All_spec} shows the spectral evolution of SN\,2020acat until the start of the nebular phase.
The initial spectrum of SN\,2020acat was obtained on 10/12/2020 ($\mathrm{MJD} = 59193.31$), approximately 1 day after the estimated explosion date. Initially, the spectra of SN\,2020acat displayed a blue excess due to the high temperature of the material, before rapidly cooling. At around $+20$ days from explosion, the blue excess had faded and the spectral line features become more dominant. The \Ha\ feature, along with \Hb\ and to a lesser degree \Hc, dominate the spectra for the first $\sim \! \! 100$ days. Helium features are also clearly present, although they are not as strong as the hydrogen ones, as well as some ionised iron features. After $\sim \! \! 100$ days, the hydrogen features have almost fully faded from the spectrum, leaving both oxygen and calcium to dominate the spectrum of SN\,2020acat, as the photosphere recedes deeper into the ejecta and the spectrum transitions into the nebular phase.

\begin{figure*}
    \centering
    \includegraphics[width=\linewidth]{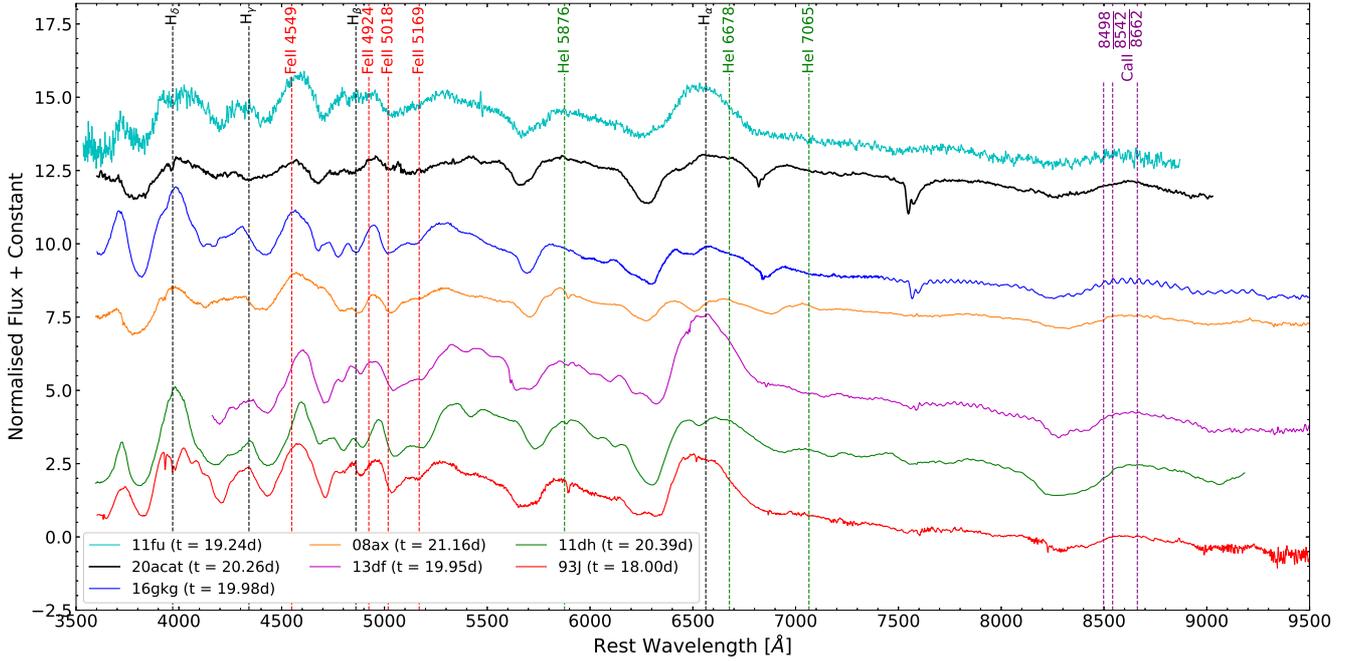}
    \caption{Spectral comparison of SN\,2020acat with several well observed SNe\,IIb roughly 20 days after explosion. Phase from explosion date for individual SNe are given in legend. All spectrum are shown in rest frame and were normalised to the peak of the \Ha\ feature. SNe are ranked in \HeI\ $\lambda 5876$ line velocity and given in rest frame. Key photospheric phase lines are marked at their rest wavelength.}
    \label{Spec_20}
\end{figure*}

\subsection{Early Phase}
\label{photospheric}
\subsubsection{Pre-maximum light}
\label{pre_max}

During the pre-maximum light phase, $\sim \! \! 0 - 20$ days after explosion, the \Ha\ feature displays a strong broad P-Cygni profile. While the hydrogen features are clearly visible throughout the photospheric phase, the \HeI\ $\lambda\lambda 5876, 6678, 7065$ lines do not produce strong features until post-max light ($\sim \! \! 25$ days after explosion), with the $5876 \, \si{\angstrom}$ being the strongest of the \HeI\ features, and the only one clearly visible in the pre-max spectra. The $6678 \, \si{\angstrom}$ feature is especially weak and seems to disappear when the \Ha\ feature is no longer visible.

In the blue region $(< 6000 \, \si{\angstrom})$ of the optical spectra, where the \FeII\ line features usually dominate, there is a strong absorption feature at $\sim \! \! 4900 \, \si{\angstrom}$ that is of similar depth as the \Hb\ feature at around $\sim \! \! 20$ days after the explosion. This feature is generally identified as \FeII\ $\lambda 5018$. However, it is generally seen with several other \FeII\ lines, all of which display a similar strength in the spectrum. This is not what is observed in the SN\,2020acat spectrum, where \FeII\ $5018$ broadly dominates over the other \FeII\ lines. Also, the $4900 \, \si{\angstrom}$ component is not broad enough to be a result of both the \FeII\ $\lambda \lambda 4924, 5018$ blending together, and there is a distinct weak line component between the \Hb\ line and the $4900 \, \si{\angstrom}$ feature that has been associated with the \FeII\ $\lambda 4924$ line. The existence of a distinct \FeII\ $\lambda 4924$ feature, along with the fact that both the $4924 \, \si{\angstrom}$ and $5169 \, \si{\angstrom}$ components are not of similar strength relative to the $5018 \, \si{\angstrom}$ line place strong doubt on the origins of the $4900 \, \si{\angstrom}$ feature as the result of solely a \FeII\ line. The question remains on what is the element that causes the $4900 \, \si{\angstrom}$ feature, within the spectra of SN\,2020acat. Multiple elements were tested by fitting the spectra at three different epochs where the $4900 \, \si{\angstrom}$ feature is visible to determine the existence of other ions that could contribute to the $4900 \, \si{\angstrom}$ feature. From this testing, the best elements that would have been able to create the $4900 \, \si{\angstrom}$ feature are either helium, nitrogen or a combination of both, enhancing the already existing \FeII\ $\lambda 5018$ line. The helium line that would result in the $4900 \, \si{\angstrom}$ feature is the \HeI\ $\lambda 5016$ line, which is significantly weaker than the \HeI\ $\lambda 5876$, with a weighted transition probability $g_{k}A_{ki}$ of $4.0116\times 10^{7}$ and $4.9496\times 10^{8}$ respectively \citep{Drake2006}, thus making it unlikely that the \HeI\ $\lambda 5016$ line is the sole responsible for the $4900 \, \si{\angstrom}$ feature. Nitrogen on the other hand has two strong lines that seem to appear within the spectrum of SN\,2020acat during the post-maximum brightness, as well as into the transition between the photospheric phase and the nebular phase. These lines being the \NII\ $\lambda\lambda\lambda 5005, 5680 \text{ and } 5942$ lines, which have equivalent transition probability to the \HeI\ $\lambda 5876$ line with a value of $g_{k}A_{ki}$ of $3.63 \times 10^{8}$ \citep{Luo_1989}, $3.47 \times 10^{8}$ \citep{doi:10.1139/p01-059} and $5.47 \times 10^{7}$ \citep{doi:10.1139/p01-059} respectively. There also exists a strong \NII\ feature at $6482 \, \si{\angstrom}$, but it overlaps with the \Ha\ feature, thus making it difficult to determine its existence at early times where the hydrogen Balmer features dominate the spectrum. While there seems to be nitrogen lines that could explain the features seen in the spectrum of SN\,2020acat it should be noted that nitrogen does not have a strong emission and as such is unlikely to strongly alter the spectrum of SN\,2020acat, unless extreme amounts of nitrogen were to be introduced. 

The pre-maximum spectrum of SN\,2020acat, taken on 18/12/2020, at around 9 days after explosion, was compared to the spectra of other five SN\,IIb at a similar epoch in Figure \ref{Spec_9}. All spectra have been corrected for reddening, given in their rest wavelength, and normalised to the peak of the \Ha\ feature. Unfortunately, SN\,1993J lacked full wavelength coverage at this epoch, covering a range of $\sim \! \! 4000-7000 \, \si{\angstrom}$. However in this phase, as the major feature at wavelengths redder than $7000\, \si{\angstrom}$ is the \CaII\ feature the missing section of SN\,1993J is not a of great concern. Relative to the other SNe\,IIb, SN\,2020acat displays broader features for all lines seen at this epoch. SN\,2020acat also displays strong hydrogen features compared to the other SNe. Unlike SN\,2008ax, SN\,2011dh and SN\,2016gkg, SN\,2020acat lacks strong iron features at around $\sim \! \! 4800 \, \si{\angstrom}$, instead showing a smooth emission from \Hb. The depth of the \CaII\ $H\&K$ feature within the spectrum of SN\,2020acat is not as deep relative to some of the other SNe, being similar to both SN\,2013df and SN\,2016gkg.

\begin{figure*}
    \centering
    \includegraphics[width=\linewidth]{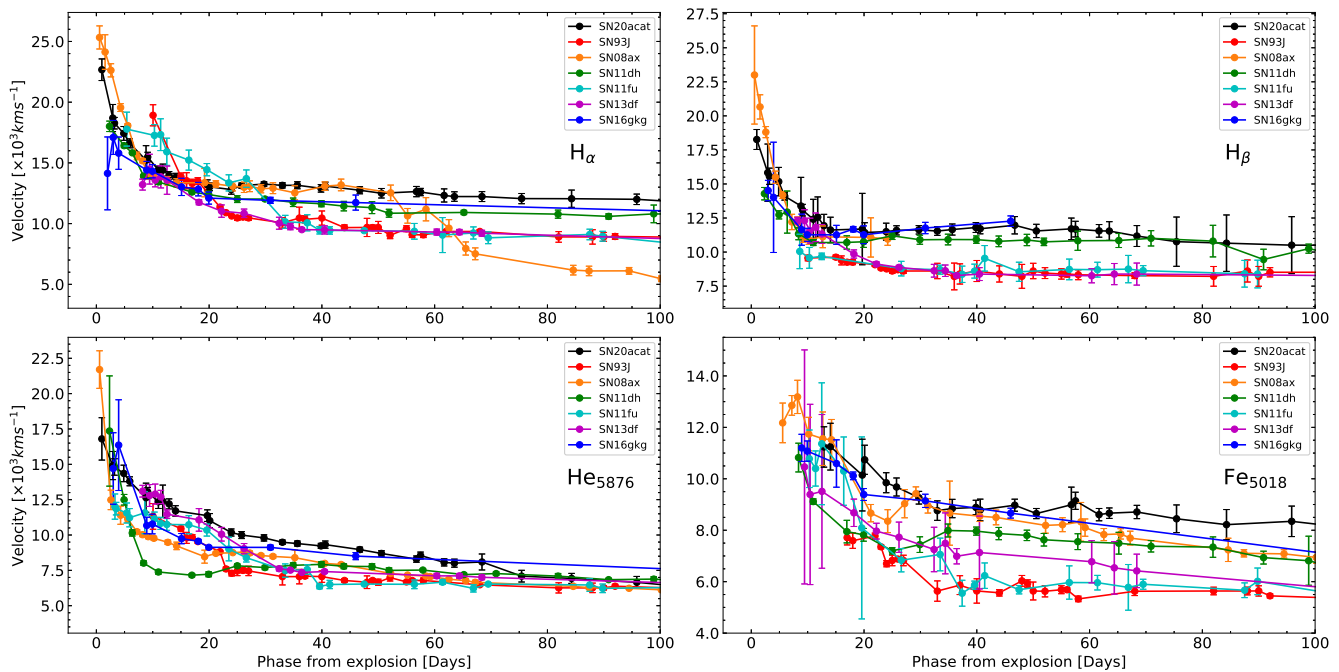}
    \caption{Line velocity evolution of \Ha, \Hb, \HeI\ $\lambda 5876$ and \FeII\ $\lambda 5018$ lines for SN\,2020acat, which are compared to those of several other SNe\,IIb, over the first 100 days from explosion. It's clearly seen that SN\,2020acat is one of the highest velocity SNe shown here, as expected from its high kinetic energy.}
    \label{line_evo_comp}
\end{figure*}

\subsubsection{Post-maximum light}
\label{post_max}
The spectrum of SN\,2020acat at $\sim \! \! 20$ days post explosion was also compared to the other SNe\,IIb at similar epochs, in Figure \ref{Spec_20}, which have been ranked by the \HeI\ $\lambda 5876$ maximum velocity. As expected, SN\,2020acat displays a very blue-shifted helium feature with only SN\,2011fu being bluer in wavelength. In all SNe spectra, the \HeI\ $\lambda 5876$ feature is strong and displays very similar shape in all SNe except SN\,1993J and SN\,2013df. The \Ha\ feature, while present in all spectra at this epoch, varies significantly in both strength and broadness among the different SNe\,IIb displayed in Figure \ref{Spec_20}. In SN\,2020acat, the \Ha\ emission is quite broad, such that the \HeI\ $\lambda 6678$ feature is visible within the emission component P-Cygni \Ha\ profile, just blue of the peak. In all spectra, the \Ha\ P-Cygni absorption feature is deeper than \HeI\ $\lambda 5876$, although in the spectrum of SN\,2008ax the features have a very similar strength. Unlike both SN\,1993J and SN\,2013df, SN\,2020acat does not display a broad flat minimum in both the \Ha\ and \HeI\ $\lambda 5876$ features, showing that they existed in distinct shells without multiple high density regions as suggested for SN\,1993J and SN\,2013df \citep{2013MNRAS.433....2S}.
All spectra also display clear NIR \CaII\ features. It should be noted that the line velocity of all major features, in the $\sim \! \! 20$ day spectrum of SN\,2020acat, are faster than those of other SNe at the same epoch, this can be clearly seen in Figure \ref{line_evo_comp}. The high velocity of all major features suggests that SN\,2020acat was a very energetic event, likely originating from a high mass progenitor star.

\subsection{Line Velocity Evolution}
\label{line_vel_text}

The expansion velocity of the ejected material in SN\,2020acat was measured by identifying the minima of the P-cygni features for the \Ha, \Hb, \HeI\ $\lambda 5876$ and \FeII\ $\lambda 5018$ lines, by fitting the absorption minimum of the line profiles with a Gaussian function. The main error associated with the line velocity originates from the noise of the individual spectrum, along with a small error associated with the fitting of the line features and the redshift correction. However, due to the high signal to noise ratio (S/N) of the spectra, the error at most epochs is not large. 
The line velocity evolution of these lines for SN\,2020acat was compared to the velocity evolution obtained from SN\,1993J, SN\,2008ax, SN\,2011dh, SN\,2011fu and SN\,2016gkg in Figure \ref{line_evo_comp}. Overall, the line velocities of elements identified in SN\,2020acat are consistently higher when compared to other SE-SNe. The \Ha\ velocity initially is rivalled by only SN\,2008ax at $\sim \! \! 22700$ \kms, being $\sim \! \! 1000$ \kms\ higher than the other SNe. The \Ha\ and \Hb\ velocities rapidly decline over the first $\sim \! \! 20$ days, before plateauing at around $\sim \! \! 13000$ \kms\ and $\sim \! \! 12000$ \kms, respectively. Both the \Ha\ and \Hb\ of SN\,2020acat display a slight increase in their velocity of $\sim \! \! 500$ \kms\ at around $\sim \! \! 30 $ days. This slight increase may result from the fitting of the spectra as there is no physical reason seen within either the spectra or light curve that would account for this increase in velocity. The \HeI\ $\lambda 5876$ feature of SN\,2020acat starts at $\sim \! \! 16800$ \kms and steadily declines for $\sim \! \! 70$ days, remaining higher than the other SE-SNe, until SN\,2020acat starts to transition into the nebular phase. The \FeII\ $\lambda 5018$ feature remains faster than the iron velocity of the other SNe throughout the spectroscopic evolution. The photospheric velocity \vph at $\sim \! \! 15.5$ days from explosion was determined to be $\sim \! \! 10000 \pm 500$ \kms. This line velocity, which corresponds with the velocity of the \FeII\ $\lambda 5169$ line, was used to break the degeneracy between the ejecta mass and kinetic energy of SN\,2020acat.

\begin{figure*}
    \centering
    \includegraphics[width=\linewidth]{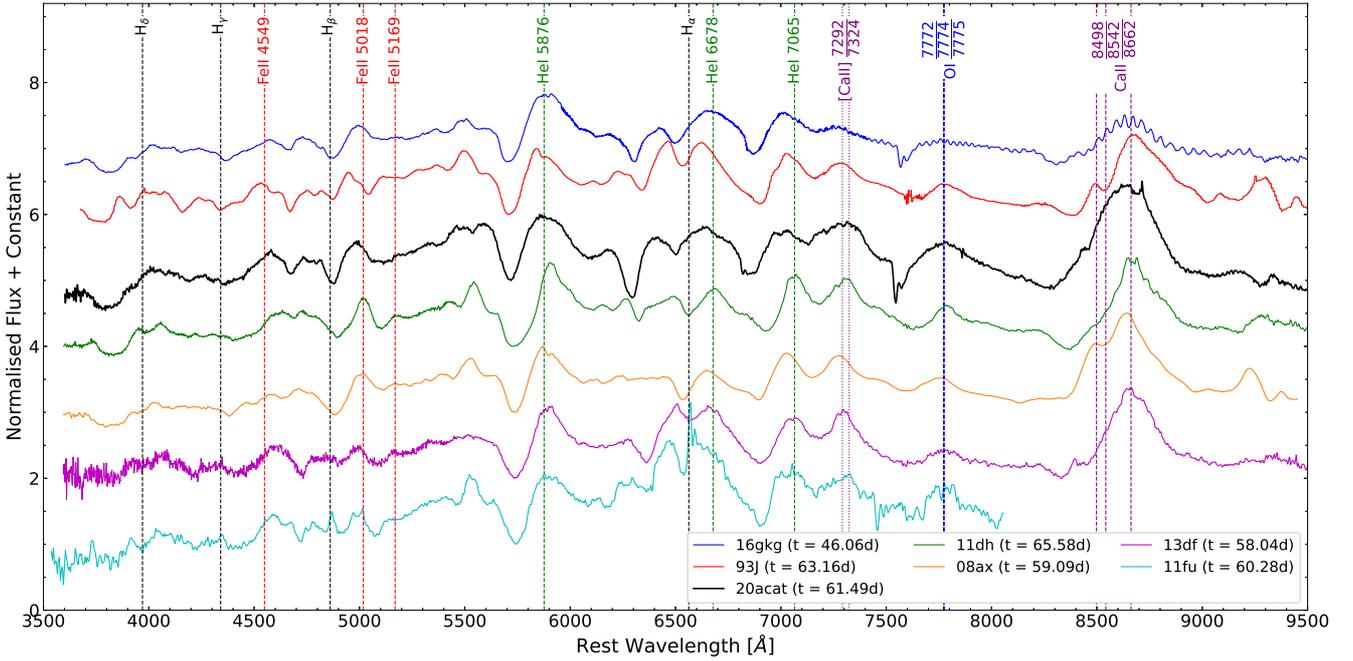}
    \caption{Spectral comparison of SN\,2020acat with several well observed SNe\,IIb roughly 60 days after their reported explosion date. All spectra are shown in rest frame and are ranked in terms of \HeI\ $\lambda 5876$ line velocity. Key transition phase element features are marked at their rest wavelength, with allowed transitions shown as the dashed lines and forbidden lines shown as dotted lines.}
    \label{Spec_60}
\end{figure*}

\subsection{Transition into the Nebular phase}
\label{transition}
Between $\sim \! \! 50-120$ days after the explosion, the spectrum of SN\,2020acat undergoes a drastic change. During this time the Balmer lines become narrower and less shallow, as the photosphere recedes deeper into the ejecta. Along with the fading of the hydrogen features, the line velocity of most elements drops at a much slower rate than during the photospheric phase, falling by $\sim \! \! 2000$ \kms\ during this period, although the \Ha\ and \Hb\ lines drop at a slower pace only decreasing by $\sim \! \! 1000$ \kms. The \HeI\ lines increase in strength relative to the \Ha\ line, and become the dominant feature as the spectrum transitions into that of a SN\,Ib. While this is happening, the NIR \CaII\ $\lambda\lambda 8498, 8542, 8662$ feature also becomes stronger, along with the weak absorption component of the \CaII\ H\&K lines. Towards the end of this phase the \Caneb\ lines $\lambda \lambda 7291, 7324$, along with the allowed \OI\ $\lambda 7773$ feature, start to appear showing the spectrum is transitioning into the nebular phase.

The spectrum of SN\,2020acat at $\sim \! \! 60$ days is also compared with those of the other SNe\,IIb at similar epochs (although the spectrum of SN\,2016gkg is $\sim \! \! 15$ days earlier than the rest due to the limited late time observations), in Figure \ref{Spec_60}. The absorption component of the \Ha\ feature in the SN\,2020acat spectrum is much deeper than those seen in the other SNe\,IIb, suggesting that a large amount of hydrogen is present at the depth of the photosphere at around $\sim \! \! 60$ days after the explosion. The presence of strong hydrogen features at this epoch may result from the progenitor having hydrogen mixed throughout the outer and parts of the inner envelope prior to collapse. \Ha\ is also still significantly broader than those seen for other SNe\,IIb.

\begin{figure*}
    \centering
    \includegraphics[width=\linewidth]{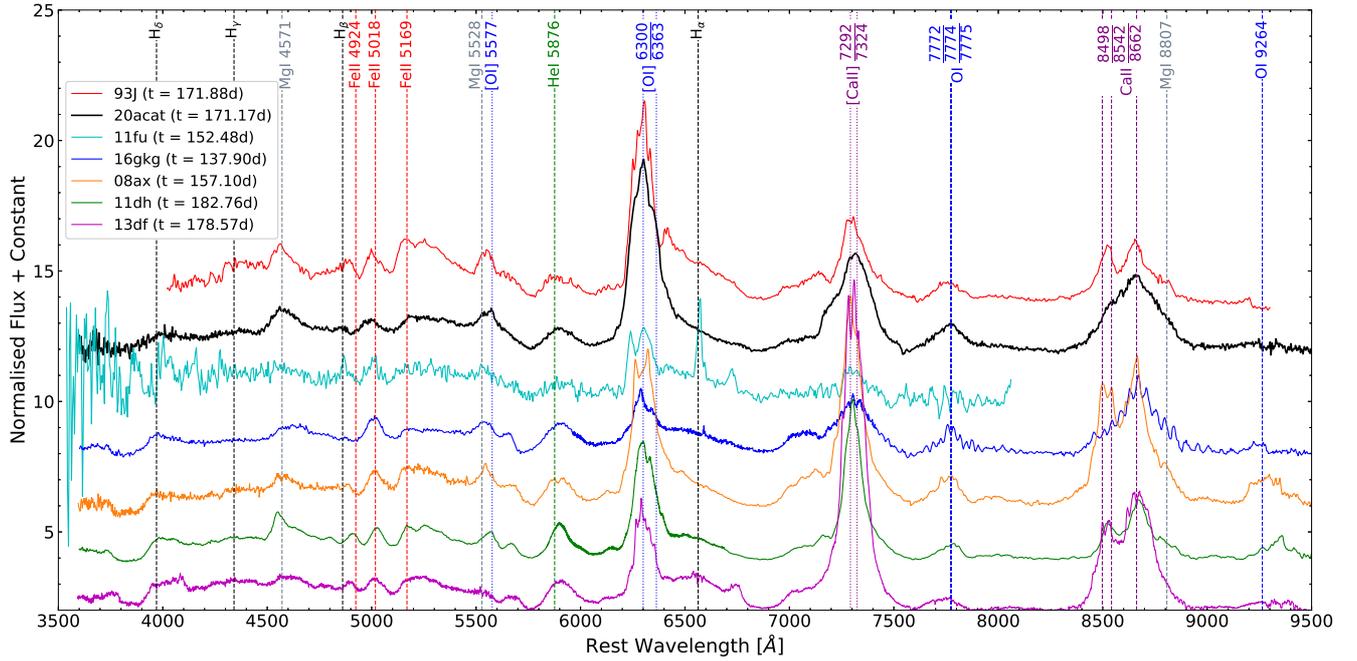}
    \caption{Spectral comparison of SN\,2020acat with several other well observed SNe\,IIb. Phase from explosion date are given in legend and all spectrum are shown in rest frame. SNe are ordered with respect to the flux ratio of the \Oneb\ and \Caneb\ nebular features. Key nebular phase line features are marked at their rest wavelengths, with allowed transitioned shown as the dashed lines and forbidden lines shown as dotted lines.}
    \label{Spec_170}
\end{figure*}

Although the \Ha\ feature remains broad in SN\,2020acat, the \HeI\ velocities have dropped to in the velocity distribution given by the other SNe\,IIb. Overall, the spectrum of SN\,2020acat at this epoch is similar to that of SN\,2016gkg, which still displays a relatively strong \Ha\ feature (although it should be noted that the spectrum of SN\,2016gkg is $\sim \! \! 15$ days earlier than that of SN\,2020acat, which may affect the similarity between the two SNe). The \Ha\ and \HeI\ $\lambda 5876$ features of all SNe are of similar widths. All SNe display strong \CaII\ features in the NIR region, with both SN\,1993J and SN\,2008ax displaying a double peaked profile. The region within the spectrum of SN\,2020acat at around $4800 - 4900 \, \si{\angstrom}$ displays both a noticeable \Hb\ and a strong feature within the \FeII\ portion of the region. This strong feature is unlikely result from only the \FeII\ lines, as it is significantly stronger than other surrounding \FeII\ line features, all of which have similar strength, and is not broad enough to be a blend of multiple \FeII\ lines. As mentioned in Section \ref{pre_max} with the early time phase spectra, both helium and iron are likely causing this strong feature, with a potential contribution from a fraction of nitrogen. 

If the $\sim \! \! 4900 \, \si{\angstrom}$ feature is indicative to the presence of nitrogen in the spectrum of SN\,2020acat, it is expected that there should be other \NII\ features should be detected, including the \NII\ $\lambda\lambda 5680, 6611$ lines. These lines can be associated with weak features seen in the $\sim \! \! 60$ day spectrum of SN\,2020acat in Figure \ref{Spec_60}, at $\sim \! \! 5500 \text{ and } 6450 \, \si{\angstrom}$, respectively. The feature potentially produced by the \NII\ $\lambda 5680$ line is located blue of the \Oneb\ $\lambda 5577$ feature, a strong emission feature in all SNe except SN\,2013df and SN\,2020acat. The potential \NII\ $6611 \, \si{\angstrom}$ feature displays a small absorption between the \Ha\ peak and the \HeI\ $\lambda 6678$ absorption minima. The $6611 \, \si{\angstrom}$ feature is significantly weaker than the $5680 \, \si{\angstrom}$ and $4900 \, \si{\angstrom}$ features, likely due to a combination of the hydrogen and helium dominating the spectrum in this region. All of these absorption features are associated with nitrogen corresponding to a line velocity of $\sim \! \! 8000 \pm 500$ \kms, suggesting that they all result from the same nitrogen-containing shell. \citet{Jerkstrand_2015} discuss models that display strong \NII\ within the late time ($> 100 $ days). However, the discovery of nitrogen in the spectrum of a SN\,IIb at this early epoch has not been discussed before. In order to determine the existence of the \NII\ lines, more detailed modelling of the spectrum at this and earlier epochs are required, which is beyond the scope of this work. 

\subsection{Nebular Phase}
\label{nebular}

At around $\sim \! \! 120$ days after the explosion of SN\,2020acat, the spectra transition into the nebular phase. During this phase, the \Ha\ feature disappears from the spectrum and the \HeI\ features decrease in strength, while both the \Oneb\ $\lambda\lambda 6300, 6363$ and the \Caneb\ $\lambda \lambda 7291, 7323$ doublets become stronger and dominate the spectrum. The spectrum of SN\,2020acat at $\sim \! \! 170$ days was compared to those of several SNe\,IIb at the same epoch, see Figure \ref{Spec_170}. Once again, the spectrum of SN\,2016gkg differs significantly in epoch compared to the other SNe. All the spectra of the SNe shown in Figure \ref{Spec_170} are dominated by either the \Oneb\ $\lambda\lambda 6300, 6363$ or the \Caneb\ $\lambda\lambda 7291, 7324$ feature at this epoch. We also identify \Mgneb\ $\lambda 4571$, \OI\ $\lambda\lambda 7772, 7774$ and the NIR \CaII. During this phase, the spectrum of SN\,2020acat appears most similar to that of SN\,1993J, both dominated by the \Oneb\ feature over the \Caneb\ feature. Both of these SNe lack a spectral feature at around $5700 \, \si{\angstrom}$, that is present in the SNe that are dominated by \Caneb. However, unlike SN\,1993J and several other SNe\,IIb, SN\,2020acat does not display a double peak in the \CaII\ NIR feature, similar to SN\,2016gkg. Another difference between SN\,2020acat and SN\,1993J is the lack of a small feature around $6400 \, \si{\angstrom}$, associated with \Ha. This suggests that hydrogen in SN\,2020acat, while mixed deep into the outer layers of the progenitor star as indicated by the strong \Ha\ visible at earlier epochs, is unlikely to penetrate deep into the inner layers. 

\begin{figure}
    \centering
    \includegraphics[width=\linewidth]{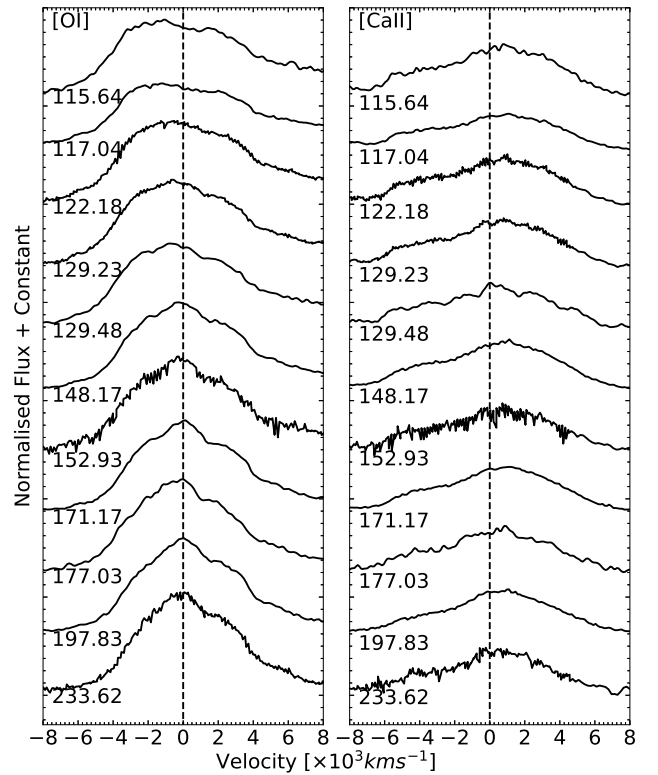}
    \caption{Line profile of \Oneb\ $\lambda\lambda 6300, 6363$ and \Caneb\ $\lambda\lambda 7292, 7324$ peaks within the late time spectra of SN\,2020acat. Dashed lines are velocities corresponding to $6300 \, \si{\angstrom}$ and $7291 \, \si{\angstrom}$ lines. The epoch from estimated explosion date for each spectrum is given in rest frame.}
    \label{O+Ca_peaks}
\end{figure}

Both the \Oneb\ and \Caneb\ features present in nebular time spectrum can be used to probe the asymmetrical nature of the ejecta, as done with SN\,2003bg by \citet{Mazzali1284}. 
The evolution of the shape of both the \Oneb\ and \Caneb\ peaks in the spectra of SN\,2020acat are shown in Figure \ref{O+Ca_peaks}. 
The spectrum at $+ 117.04 $ days still displays a remnant of \Ha, seen by the flat-topped profile that would not be present if only \Oneb\ was emitting in that region.
Once the \Ha\ feature has fully faded from the spectrum, at $\sim \! \! 130$ days, the \Oneb\ feature displays a strong symmetric shape. When compared with the \Oneb\ features of other SE-SNe, the \Oneb\ of SN\,2020acat does not display a strong double peak feature, and appears more similar to both SN\,2011fu and SN\,2013df. Instead, the nebular spectra of SN\,2020acat display a small bump on the red side of the peak, which is due to the \Oneb $\lambda 6363$ line. The centroid of the \Oneb\ peak is aligned with the $6300 \, \si{\angstrom}$, while the centroid of the \Caneb\ peak is slightly shifted by $\sim \! \! 1000$ \kms, which seems to move towards zero velocity as the spectra evolve. The \Caneb\ feature is broader than those of other SNe\,IIb by several thousand \kms, likely a result of the high explosion energy causing a large dispersion in the calcium velocity distribution.

\subsection{Oxygen mass and \Caneb/\Oneb\ ratio}
\citet{1989ApJ...343..323F} suggested that the ratio between the flux of the \Caneb\ $\lambda \lambda 7291, 7324$ and the \Oneb\ $\lambda \lambda 6300, 6363$ lines provides a good way of estimating whether the progenitor had a large or small main-sequence mass. This arises from the assumption that the flux of the oxygen emission region is directly related to the mass of oxygen formed throughout the evolution of the progenitor star, while the flux of the calcium emitting region is dependent solely on the mass of calcium synthesised during the explosion, and not effected by the mass of the progenitor during its life cycle. Therefore, a large ratio, \Caneb / \Oneb $ \gtrsim \sim \! \! 1$, is expected to be the result of the progenitor having a small main-sequence mass. It was shown by \citet{refId0} that the \Caneb\ / \Oneb\ ratio within the spectrum at late enough times ($> 150$ days) is expected to stay stable over very long periods. For SN\,2020acat, at $\sim \! \! 170$ days the \Caneb/\Oneb\, ratio was found to be $\sim \! \! 0.5$, similar to that seen for SN\,1993J and SN\,2011fu at similar epochs, which suggests that the progenitor of SN\,2020acat had a large \mzams. However, it should be noted that this method is not a robust tool for obtaining \mzams. The ratio between the \Caneb\ $\lambda\lambda 7292, 7324$ and \Oneb\ $\lambda\lambda 6300, 6363$ varies strongly among the different SNe\,IIb displayed in Figure \ref{Spec_170}, with SN\,2008ax, SN\,2011dh and SN\,2013df all with a stronger forbidden calcium feature relative to the oxygen one. SN\,1993J, SN\,2011fu, SN\,2016gkg and SN\,2020acat all display a stronger oxygen feature which dominates over the calcium, although in SN\,2016gkg the peaks of the oxygen and calcium features are very similar at $\sim \! \! 140$ days.
It should be noted that the SN\,2016gkg spectrum is $\sim \! \! 30$ days earlier than the other SNe, and the ratio of peaks may have changed by quite a bit by $\sim \! \! 170$ days, especially given the proximity to unity that the \Caneb/\Oneb\ ratio had at $\sim \! \! 140$ days.

From the nebular phase spectra of SN\,2020acat, an estimation of the oxygen mass can be made, which can provide insights into the expected progenitor mass. The relationship between the observed \Oneb\ emission peak and the mass of oxygen was described by \citet{1986ApJ...310L..35U}, which is expected to hold within a high density limit $(N_{e} \geq 10^{6} \mathrm{cm^{-3}})$, and is given by:

\begin{ceqn}
\begin{align}    
    M_\mathrm{O} = 10^{8} F([\mathrm{O_{I}}]) D^{2} \times exp\left(\frac{2.28}{T_{4}}\right), 
    \label{M_oxy}
\end{align}
\end{ceqn}

where $M_\mathrm{O}$ is the mass of neutral oxygen in \msun, F(\Oneb) is the flux of the \Oneb\ $\lambda\lambda 6300, 6363$ peak in $\mathrm{erg s^{-1} cm^{-2}}$, $D$ is the distance in Mpc and $T_{4}$ is the temperature of the oxygen emitting region of the spectrum in units of $10^{4} K$. The temperature of the \Oneb\ region can be determined by the ratio of fluxes between \Oneb\ $5577 \, \si{\angstrom}$ and the \Oneb\ $\lambda\lambda 6300, 6363$ lines. However, determining the flux of the \Oneb\ $5577 \, \si{\angstrom}$ peak is not easy, as it can blend with \FeII\ lines, distorting the flux value of the peak. As such, a temperature of $T_{4} = 0.4 K$ was used, which arises from the assumption that within the oxygen emitting region during the nebular phase the density is high and the temperature is low \citep{refId0}. Using this low temperature, along with a F(\Oneb) $= 8.41 \times 10^{-14}$ \flux\ derived from the spectrum taken at $\sim \! \! 170$ days post explosion, results in an oxygen mass of $M_\mathrm{O} = 3.13 \pm 0.07$ \msun. This is a large oxygen mass given the ejecta mass derived from the light curve. However, realistically the flux of oxygen is not constant over time as the SN fades and the spectra transitions further into the nebular phase. As such, different oxygen masses can be obtained as the spectra evolve, the values of which are given in Table \ref{O_masses}. Due to the changing flux as SN\,2020acat fades, the value of $M_\mathrm{O}$ derived from the spectrum taken $\sim \! \! 171$ days after the explosion can be considered an upper limit to the oxygen mass.

\begin{table}
    \centering
    \begin{tabular}{ccc}
    \hline
    Phase & Flux & $M_\mathrm{O}$ \\
    $[days]$ & $[\times 10^{-14} \mathrm{erg \, s^{-1} \, cm^{-2}}]$ & [\msun] \\
    \hline
    171.17 & 8.407 & 3.13 (0.07) \\
    177.02 & 7.406 & 2.76 (0.05) \\
    197.83 & 7.691 & 2.87 (0.06) \\
    233.62 & 2.590 & 0.97 (0.02) \\
    \hline
    \end{tabular}
    \caption{Mass of oxygen obtained from the late time spectra of SN\,2020acat using equation. \ref{M_oxy}. The flux from the spectrum taken 171.17 days from explosion can be used for an upper limit to the mass of oxygen. The error associated with each epoch is given in the parenthesis.}
    \label{O_masses}
\end{table}

\begin{figure}
    \centering
    \includegraphics[width=\columnwidth]{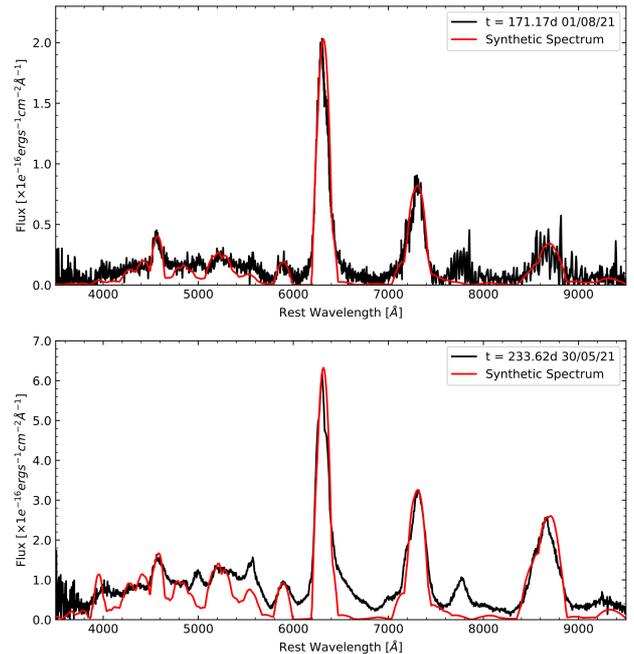}
    \caption{Plot of the nebular phase spectra (black) of SN\,2020acat, taken at $\sim \! \! 171.17 \text{ (top) and } 233.62$ (bottom) days post explosion, along with the one-zone nebular model of \citet{Mazzali_2001}.}
    \label{Dual_Neb_Models}
\end{figure}

Two of the nebular spectra of SN\,2020acat ($t = 171.17 \text{ and } t = 233.63$) were modelled using our SN nebular spectrum synthesis code \citep[e.g.,][]{mazzali2007_02ap}. Briefly, the code computes the emission of gamma rays and positrons by the radioactive decay of \Nifs\, and \Cofs, and computes their deposition in the expanding SN ejecta, using a gamma-ray opacity $\kappa_{gamma} = 0.027 {\rm g cm}^{-2}$ and a positron opacity $\kappa_{e^+} = 7 {\rm g cm}^{-2}$. Following the prescriptions of \citet{axelrod80}, the deposited energy is used to heat the gas via collisional processes. Heating is then balanced by cooling via radiation in mostly forbidden lines. Ionization and recombination rates are balanced and the level populations within different ions is computed in non-local thermodynamic equilibrium (NLTE). The SN nebula is assumed to be optically thin at late times and radiation transport is not performed.
For SN\,2020acat we use a simple one-zone version of the code, which allows us to determine the basic parameters of the inner ejecta without making assumptions about the density distribution. Clearly, because of such an approach, our results must be regarded as approximate.
We assumed for the two spectra at $\sim \! \! 231 \text{ and } 170$ days after explosion, used a distance modulus $\mu = 32.74$\,mag and a total reddening $E(B-V) = 0.0207$ mag. The outer boundary velocity for the part of the nebula that contributes to the emission was set at 5100\,\kms, based on the width of the emission lines. Given that we did not use a stratified model of the ejecta, we should expect that deposition efficiency decreases with time faster than in the real SN. This means that at later epochs somewhat larger values of the masses are required to fit the spectrum \citep{Mazzali_2001}. 
Figure \ref{Dual_Neb_Models} shows the two observed spectra and the corresponding synthetic spectra. The main emission features are [\OI] 6300,6363\,\AA, \MgI] 4570\,\AA, and \CaII] 7291,7324\,\AA. Weaker lines include [\OI] 5577\,\AA, which is sensitive to recombination, the [\OI] recombination line near 7773\,\AA, which is not reproduced in our model as we do not consider recombination emission, \NaI\,D, and several [\FeII] lines, mostly near 5200\,\AA, which are important to determine the abundance of \Nifs. 
The mean masses of the elements that contribute to the spectra are \Nifs\, $= 0.10 \pm 0.02$ \msun, $M_\mathrm{O} = 1.0 \pm 0.10$ \msun, $M_\mathrm{C} = 0.20 \pm 0.05$ \msun, $M_\mathrm{Ca} = 0.045 \pm 0.005$ \msun, $M_\mathrm{Mg} = 0.0012 \pm 0.0004$ \msun, $M_\mathrm{Na} = 0.0008 \pm 0.0001$ \msun.  Small amounts of Si and S were also included for consistency, but these elements do not produce strong lines in the optical range.  The spectra do not appear to be fully nebular. In particular, a feature near 5000\,\AA\ may still have a P-Cygni profile from \FeII\ multiplet 42 lines. 
The ejected mass within the boundary velocity is $1.5 \pm 0.15$\,\msun. This indicates a moderately massive CO core ($\sim \! \! 3$\,\msun\ if we assume a neutron star remnant), and is consistent with previous results for stripped-envelope SNe \citep[e.g., ][Figure 14]{mazzali2021}, indicating a progenitor zero-age main sequence mass of $\sim \! \! 20$ \msun.

\section{Conclusions}
\label{Conclusion}

SN\,2020acat is a well observed SN\,IIb, with a highly constrained explosion date, that was thorough observed in the UV to NIR photometric regions during the peak time, and extensive optical follow-up during the transition into the nebular phase. Spectroscopically SN\,2020acat was followed for $\sim \! \! 230$ days resulting in a comprehensive optical campaign from pre-maximum light to the start of the nebular phase. The follow-up campaign was unfortunately halted as SN\,2020acat moved into Solar conjunction. The photometric and spectroscopic data set gives SN\,2020acat one of the best follow-up campaign available for a SN\,IIb caught within a few days of its explosion. 

SN\,2020acat displays a very fast initial rise, reaching a peak of the pseudo-bolometric light curve in $\sim \! \! 15$ days of the estimated explosion date, more rapidly than other SNe\,IIb. The fast rise in the pseudo-bolometric light curve, and the underestimation of the light curve model used in Section \ref{bol_sec}, may result from the \Nifs\ being mixed into the outer layers of the ejecta. More detailed modelling of the pseudo-bolometric light curve is required to determine the distribution of \Nifs\ within the ejecta of SN\,2020acat. Along with the fast rise of the pseudo-bolometric light curve SN\,2020acat lacks any visible decline in its early time UV and optical light curves, normally associated with the shock-cooling tail. While it might be possible that the shock-breakout phase was completely missed by observations, the dim initial observation, along with both the shape of the rising light curves and the tight constraint on the estimated explosion date, suggest that SN\,2020acat lacked any evident shock-cooling phase. The lack of an extended shock-cooling phase strongly implies that the progenitor of SN\,2020acat was a compact object that lacked an extended hydrogen envelope, as was seen with the light curves of SN\,2008ax.

Spectroscopically, SN\,2020acat shows strong hydrogen features throughout its evolution into the nebular phase, covering the first $\sim \! \! 100$ days, at which point oxygen, along with calcium, starts to dominate the spectrum. The clear presence of hydrogen well into the transition to the nebular phase suggests that SN\,2020acat possessed either a very dense thin hydrogen envelope prior to explosion or the hydrogen was mixed deep into the outer layers through some means of convection.
The spectral feature usually associated with the \FeII\ $\lambda 5018$ line was seen to be much stronger than the surrounding \FeII\ lines, suggesting it was likely a blend with other species.
From analysis of the spectra at multiple epochs around peak time, it was determined that the feature was likely enhanced by the presence of a combination of helium and nitrogen within the ejecta. While the helium line that could have enhanced this feature has been reported prior, the presence of nitrogen is not expected in the spectra of SNe\,IIb. One possible source for the potential nitrogen is if the progenitor of SN\,2020acat was a massive enough star, then some nitrogen may still remain from the helium burning stage and appears in the spectrum as weak \NII\ features. Future work on modelling the evolution of SN\,2020acat is required to investigate the potential presence of nitrogen in its spectra at early epochs.

\section*{Acknowledgements}

    This work makes use of observations from the Las Cumbres Observatory network. The LCO team is supported by NSF grants AST-1911225 and AST-1911151, and NASA \textit{SWIFT} grant 80NSSC19K1639. Based in part on observations made with the Liverpool Telescope operated on the island of La Palma by Liverpool John Moores University in the Spanish Observatorio del Roque de los Muchachos of the Institutode Astrofisica de Canarias with financial support from the UK Science and Tech-nology Facilities Council. This work is partially based on observations collected at the Copernico 1.82 m and Schmidt 67/92 Telescopes operated by INAF Osservatorio Astronomico di Padova at Asiago, Italy. Based on observations made with the Nordic Optical Telescope, owned in collaboration by the University of Turku and Aarhus University, and operated jointly by Aarhus University, the University of Turku and the University of Oslo, representing Denmark, Finland and Norway, the University of Iceland and Stockholm University at the Observatorio del Roque de los Muchachos, La Palma, Spain, of the Instituto de Astrofisica de Canarias. The data presented here were obtained in part with ALFOSC, which is provided by the Instituto de Astrofisica de Andalucia (IAA) under a joint agreement with the University of Copenhagen and NOT. This work is partly based on the NUTS2 programme carried out at the NOT. NUTS2 is funded in part by the Instrument Center for Danish Astrophysics (IDA). CA is supported by NASA grant 80NSSC19K1717 and NSF grants AST-1920392 and AST-1911074. TM-B. acknowledges financial support from the Spanish Ministerio de Ciencia e Innovaci\'on (MCIN), the Agencia Estatal de Investigaci\'on (AEI) 10.13039/501100011033 under the PID2020-115253GA-I00 HOSTFLOWS project, and from Centro Superior de Investigaciones Cient\'ificas (CSIC) under the PIE project 20215AT016. AR acknowledges support from ANID BECAS/DOCTORADO NACIONAL 21202412. LG acknowledges financial support from the Spanish Ministerio de Ciencia e Innovaci\'on (MCIN), the Agencia Estatal de Investigaci\'on (AEI) 10.13039/501100011033, and the European Social Fund (ESF) "Investing in your future" under the 2019 Ram\'on y Cajal program RYC2019-027683-I and the PID2020-115253GA-I00 HOSTFLOWS project, and from Centro Superior de Investigaciones Cient\'ificas (CSIC) under the PIE project 20215AT016. IA is a CIFAR Azrieli Global Scholar in the Gravity and the Extreme Universe Program and acknowledges support from that program, from the European Research Council (ERC) under the European Union's Horizon 2020 research and innovation program (grant agreement number 852097), from the Israel Science Foundation (grant number 2752/19), from the United States - Israel Binational Science Foundation (BSF), and from the Israeli Council for Higher Education Alon Fellowship. MG is supported by the EU Horizon 2020 research and innovation programme under grant agreement No 101004719. MN is supported by the European Research Council (ERC) under the European Union's Horizon 2020 research and innovation programme (grant agreement No.~948381) and by a Fellowship from the Alan Turing Institute. N.E.R. acknowledges partial support from MIUR, PRIN 2017 (grant 20179ZF5KS), from the Spanish MICINN grant PID2019-108709GB-I00 and FEDER funds and by the program Unidad de Excelencia Mar\'ia de Maeztu CEX2020-001058-M. TMR acknowledges the financial support of the Vilho, Yrj{\"o} and Kalle V{\"a}is{\"a}l{\"a} Foundation of the Finnish academy of Science and Letters. Y.-ZC is funded by China Postdoctoral Science Foundation (grant no. 2021M691821). PL and ME acknowledge support from the Swedish Research Council. P.C is supported by a research grant (19054) from VILLUM FONDEN.


\section*{Data Availability}
Spectroscopic data will be made available on the Weizmann Interactive Supernova Data Repository (WISeREP) at https://wiserep.weizmann.ac.il/. Photometric data is available as suplimentary data on the online version of this work.



\bibliographystyle{mnras}
\bibliography{References.bib} 

\section*{Affiliations}
\footnotesize{
$^{1}$Astrophysical Research Institute Liverpool John Moores University, Liverpool L3 5RF, UK \\
$^{2}$Max-Planck Institute for Astrophysics, Karl-Schwarzschild-Str. 1, D-85748 Garching, Germany \\
$^{3}$Institute for Astronomy, University of Hawai'i at Manoa, 2680 Woodlawn Dr., Hawai'i, HI 96822, USA \\
$^{4}$European Southern Observatory, Alonso de C\'ordova 3107, Casilla 19, Santiago, Chile \\
$^{5}$The School of Physics and Astronomy, Tel Aviv University, Tel Aviv 69978, Israel \\
$^{6}$CIFAR Azrieli Global Scholars program, CIFAR, Toronto, Canada \\
$^{7}$INAF – Osservatorio Astronomico di Padova, Vicolo dell'Osservatorio 5, I-35122 Padova, Italy \\
$^{8}$DIRAC Institute, Department of Astronomy, University of Washington, 3910 15th Avenue NE, Seattle, WA 98195, USA) \\
$^{9}$Las Cumbres Observatory, 6740 Cortona Dr, Suite 102, Goleta, CA 93117-5575, USA \\
$^{10}$Department of Physics, University of California, Santa Barbara, CA 93106-9530, USA \\
$^{11}$Physics Department and Tsinghua Center for Astrophysics (THCA), Tsinghua University, Beijing, 100084, China \\
$^{12}$DTU Space, National Space Institute, Technical University of Denmark, Elektrovej 327, DK-2800 Kgs. Lyngby, Denmark \\
$^{13}$Institute of Space Sciences (ICE, CSIC), Campus UAB, Carrer de Can Magrans s/n, 08193 Barcelona, Spain \\
$^{14}$Department of Astronomy, AlbaNova University Center, Stockholm University, SE-10691 Stockholm, Sweden \\
$^{15}$The Oskar Klein Centre, AlbaNova, SE-10691 Stockholm, Sweden \\
$^{16}$Institut d'Estudis Espacials de Catalunya (IEEC), E-08034 Barcelona, Spain \\
$^{17}$Astronomical Observatory, University of Warsaw, Al. Ujazdowskie 4, 00-478 Warszawa, Poland \\
$^{18}$Center for Astrophysics $|$ Harvard \& Smithsonian, 60 Garden Street, Cambridge, MA 02138-1516, USA \\
$^{19}$The NSF AI Institute for Artificial Intelligence and Fundamental Interactions \\
$^{20}$Cardiff Hub for Astrophysics Research and Technology, School of Physics \& Astronomy, Cardiff University, Queens Buildings, The Parade, Cardiff, CF24 3AA, UK \\
$^{21}$Birmingham Institute for Gravitational Wave Astronomy and School of Physics and Astronomy, University of Birmingham, Birmingham B15 2TT, UK \\
$^{22}$IAASARS, Observatory of Athens, 15236, Penteli, Greece \\
$^{23}$Nordic Optical Telescope, Apartado 474, 38700 Santa Cruz de La Palma, Santa Cruz de Tenerife, Spain \\
$^{24}$Facultad de Ciencias Astron\'omicas y Geof\'isicas, Universidad Nacional de La Plata, Paseo del Bosque S/N, B1900FWA, La Plata, Argentina \\
$^{25}$Departamento de Ciencias F\'{i}sicas – Universidad Andres Bello, Avda. Rep\'{u}blica 252, Santiago, Chile \\
$^{26}$Millennium Institute of Astrophysics, Nuncio Monsenor S\'{o}tero Sanz 100, Providencia, Santiago, Chile \\
$^{27}$The Cosmic Dawn Center (DAWN), Niels Bohr Institute, University of Copenhagen, Rådmandsgade 62, 2200 København N, Denmark \\
$^{28}$Inter-University Centre for Astronomy and Astrophysics, Pune - 411007, India \\
$^{29}$Astrophysics Research Centre, School of Mathematics and Physics, Queen's University Belfast, Belfast BT7 1NN, UK}






\bsp	
\label{lastpage}
\end{document}